\documentclass[12pt,twoside]{article}   %For printing on two sides of page
\usepackage[super,sort,comma]{natbib}
\usepackage{fancyhdr}		%Gives headers and footers defined below
\usepackage[section]{placeins}   %
\usepackage{graphicx}
\usepackage{amsmath,amssymb,amsfonts}
\usepackage{algorithmicx}
\usepackage{textcomp}
\usepackage{algorithm}
\usepackage{algpseudocode}

\makeatletter \renewcommand\@biblabel[1]{$^{#1}$} \makeatother
\newcommand{\note}[1]{\mbox{}\\ \noindent \rule{16cm}{0.5mm} \\
{\em #1} \\ \noindent \rule{16cm}{0.5mm}
\typeout{    }
\typeout{***********note active on this page *************************}
\typeout{Note: #1  }
\typeout{****************************************end Note}
}

\renewcommand{\note}[1]{}

\newcommand{\cen}[1]{\begin{center} #1 \end{center}}

\usepackage{hyperref}
\hypersetup{ colorlinks,
    citecolor=blue,
    filecolor=blue,
    linkcolor=blue,
    urlcolor=blue
}

\usepackage{xcolor}
\definecolor{gray}{rgb}{0.6,0.6,0.6}
\definecolor{red}{rgb}{0.85,0,0}
\definecolor{green}{rgb}{0,0.85,0}
\definecolor{blue}{rgb}{0,0,0.85}
\definecolor{beige}{rgb}{0.92,0.87,0.78}
\usepackage[all]{hypcap}    %causes link to figures to go to figure, not caption
\begin{document}

\cen{\sf {\Large {\bfseries Report on the AAPM deep-learning spectral CT Grand Challenge} \\  
%\vspace*{10mm}
\vspace*{5mm}
Emil Y. Sidky and Xiaochuan Pan
 } \\
Department of Radiology, The University of Chicago, 5841 S. Maryland Ave., Chicago, IL 60637, USA
%\vspace{5mm}\\
\vspace{3mm}\\
Version typeset \today\\
}

\pagenumbering{roman}
\setcounter{page}{1}
\pagestyle{plain}
Author to whom correspondence should be addressed. email: sidky@uchicago.edu\\
% note, probably best not to use a student's e-mail as it won't be valid for
% very long.

\begin{abstract}
\noindent
{\bf Background:}
This Special Report summarizes the 2022 AAPM Grand Challenge on Deep-Learning spectral Computed Tomography (DL-spectral CT)
image reconstruction.\\
\noindent
{\bf Purpose:}
The purpose of the challenge is to develop the most accurate image reconstruction algorithm
for solving the inverse problem associated with a fast kVp switching dual-energy CT scan using a three tissue-map
decomposition. Participants could choose to use deep-learning (DL), iterative, or a hybrid approach.\\
{\bf Methods:}
The challenge is based on a 2D breast CT simulation, where the simulated breast phantom consists
of three tissue maps: adipose, fibroglandular, and calcification distributions. The phantom specification
is stochastic so that multiple realizations can be generated for deep-learning approaches.
A dual-energy scan is simulated where the X-ray source potential of successive views alternates between  50 and 80 kilovolt-peak (kVp).
A total of 512 views are generated, yielding 256 views for each source voltage.
We generate 50 and 80 kVp images by use of filtered back-projection (FBP) on negative logarithm processed
transmission data. For participants who develop a DL approach, 1000 cases are available.
Each case consists of the three 512x512 tissue maps, 50 and 80 kVp transmission data sets and their corresponding
FBP images. The goal of the DL network would then be to predict the material maps from either the transmission
data, FBP images, or a combination of the two.
For participants developing a physics-based approach, all the modeling parameters are made available:
geometry, spectra, and tissue attenuation curves.
The provided information also allows for hybrid approaches where physics is exploited as well
as information about the scanned object derived from the 1000 training cases.
Final testing is performed by computation of root-mean-square-error (RMSE) for predictions on the tissue
maps from 100 new cases.\\
{\bf Results:} 
Test phase submission were received from 18 research groups.
Both the winning and second place teams had highly accurate results where the RMSE was nearly zero
to single floating point precision. Results from the top ten also achieved a high degree of accuracy;
and as a result this special report outlines the methodology developed by each of these groups.\\
{\bf Conclusions:} 
The DL-spectral CT challenge successfully established a forum for developing image reconstruction algorithms
based on deep-learning that address an important inverse problem relevant for spectral CT.\\
\end{abstract}
%\note{This is a sample note.}

\newpage     %may or may not be needed

%The table of contents is for drafting and refereeing purposes only. Note
%that all links to references, tables and figures can be clicked on and
%returned to calling point using cmd[ on a Mac using Preview or some
%equivalent on PCs (see View - go to on whatever reader).
\tableofcontents

\newpage

\setlength{\baselineskip}{0.7cm}      %double spacing		

\pagenumbering{arabic}
\setcounter{page}{1}
\pagestyle{fancy}

\section{Introduction}
\label{sec:introduction}

This Special Report summarizes 
the American Association of Physics in Medicine (AAPM) sponsored 2022 Grand Challenge on deep-learning (DL) for
spectral computed tomography (CT) image reconstruction, DL-spectral CT.  The DL-spectral CT Grand Challenge is
motivated by the interest in addressing inverse problems in imaging using deep-learning, and DL-spectral CT follows
and builds off of
the 2021 AAPM Grand Challenge on deep-learning for sparse-view CT image reconstruction, DL-sparse-view CT \cite{sparseChallengeReport}.
While deep-learning has been heavily investigated for image reconstruction/restoration problems \cite{ongie2020deep}, there
has not been much work reported on the accuracy of DL methodology for inversion of the imaging models that underpin
tomographic image reconstruction. These challenges provide a forum in which to test DL methodology on inverse problems
relevant to CT image reconstruction.

Before explaining the design of the 2022 DL-spectral CT challenge, we briefly recapitulate the design and results
of the 2021 DL-sparse-view CT challenge. The sparse-view challenge was based on our paper with Lorente and Brankov \cite{sidky2020cnns},
where we attempted to use previously published work on convolutional neural networks (CNNs) to solve a sparse-view CT
image reconstruction problem. In this article we were not able to show numerical solution of the sparse-view CT solution.
This result motivated us to create the 2021 DL sparse-view CT challenge.

The sparse-view CT inverse problem  is known to have a solution by using iterative image reconstruction exploiting
gradient sparsity as demonstrated in Sidky {\it et al.}\cite{sidky2020cnns}. Because there was a published solution
to this sparse-view problem, the participants of the DL-sparse-view challenge did not have full knowledge of the projection
model as this would have allowed them to simply reconstruct the images using the gradient-sparsity exploiting algorithm.
The participants had to rely on the 4000-case training set to devise an accurate DL-based image reconstruction algorithm.
In last year's challenge, 25 competing teams submitted their predictions for the testing set, which consisted of 100 cases.
The winning team, Robust-and-stable, achieved highly accurate image reconstruction on each of the test cases and the numerical
error was small enough that it was not visible within the display gray-scale window appropriate for the test phantom.
The Robust-and-stable team published their methodology in the Proceedings of the 39th International Conference on Machine
Learning \cite{genzel2022near} and they have also made code available.

For the 2022 DL-spectral CT challenge, we constructed a simulation based on a hypothetical dual-energy breast CT
system. The goal is to recover three tissue maps, adipose, fibro-glandular, and calcification distributions, from
simulated noiseless dual-energy transmission data. Unlike the sparse-view problem of the previous challenge,
this particular inverse problem does not have a published solution,
and accordingly we provided complete knowledge of the forward model. A training set of 1000 cases for this challenge
enabled DL approaches. With the forward model available, participants could also develop iterative algorithms
that solve the physical imaging model; or hybrid physics-based and DL-based approaches could be devised.
Because the provided data is generated from the ground truth images without any inconsistencies, it is in principle
possible to recover the exact ground truth tissue maps. Accordingly, the metric used for rating performance
is the root mean square error (RMSE) of the predicted test case tissue maps from the ground truth. Because
this particular inverse problem has not been solved, we did not know in advance if the RMSE could in fact
be driven to zero.

A total of 18 teams submitted their predictions for the 100 cases of the test phase for DL-spectral CT.
As with last year's challenge, the participating teams came up with highly innovative algorithms that included
DL-based, physics-based, and hybrid approaches. 
There was, again, a good spread in RMSE scores and the winning/runner-up teams achieved accurate tissue map recovery
up to six/five decimal places. Because the accuracy level of the results was high and the methodology so varied,
we report briefly on the methodology of the top ten performing teams.

The details of the DL-spectral CT challenge are presented in Sec. \ref{sec:methods}, where the physical imaging
model and logistics of the DL-spectral CT challenge are discussed.
The results of the challenge are shown in Sec. \ref{sec:results} along with overview
descriptions of the top ten algorithms.
Finally, a discussion that summarizes what was learned about the three tissue-map recovery problem
in dual-energy CT is given in Section \ref{sec:discussion}.

\section{Methods}
\label{sec:methods}

As with DL sparse-view CT, the present DL spectral CT challenge is focused on solving an inverse problem
relevant for CT imaging; see Sec. IIA of the DL sparse-view CT challenge report for background on Inverse
Problem theory \cite{sparseChallengeReport}. The basic idea is to set up a simulation where the data are
generated from a physical model of the scanner using computer simulated images, and then to see if the
test images can be recovered accurately from perfect knowledge of the data, i.e. no noise. For DL-spectral
CT, we provide the complete physical model for the dual-energy fast kVp-switching scan configuration.
The breast phantom is composed of three tissue maps modeling adipose, fibroglandular, and calcification
distributions. For participants interested in developing a DL-based algorithm, we also provide
a training set of 1000 cases, where each case consists of a breast phantom realization, dual-energy
CT transmission data, and filtered back-projection (FBP) reconstructed low and high kVp images. Participants aim to estimate
the three tissue maps based on the provided transmission data and/or the kVp images. In this section
we cover the details of the spectral CT model, the evaluation of the results, and the logistics of the challenge.

%\subsection{Fast kVp-switching dual-energy CT simulation and the breast phantom}
\subsection{Fast kVp-switching dual-energy CT simulation}

\begin{figure}[!t]
\centerline{\includegraphics[width=0.6\columnwidth]{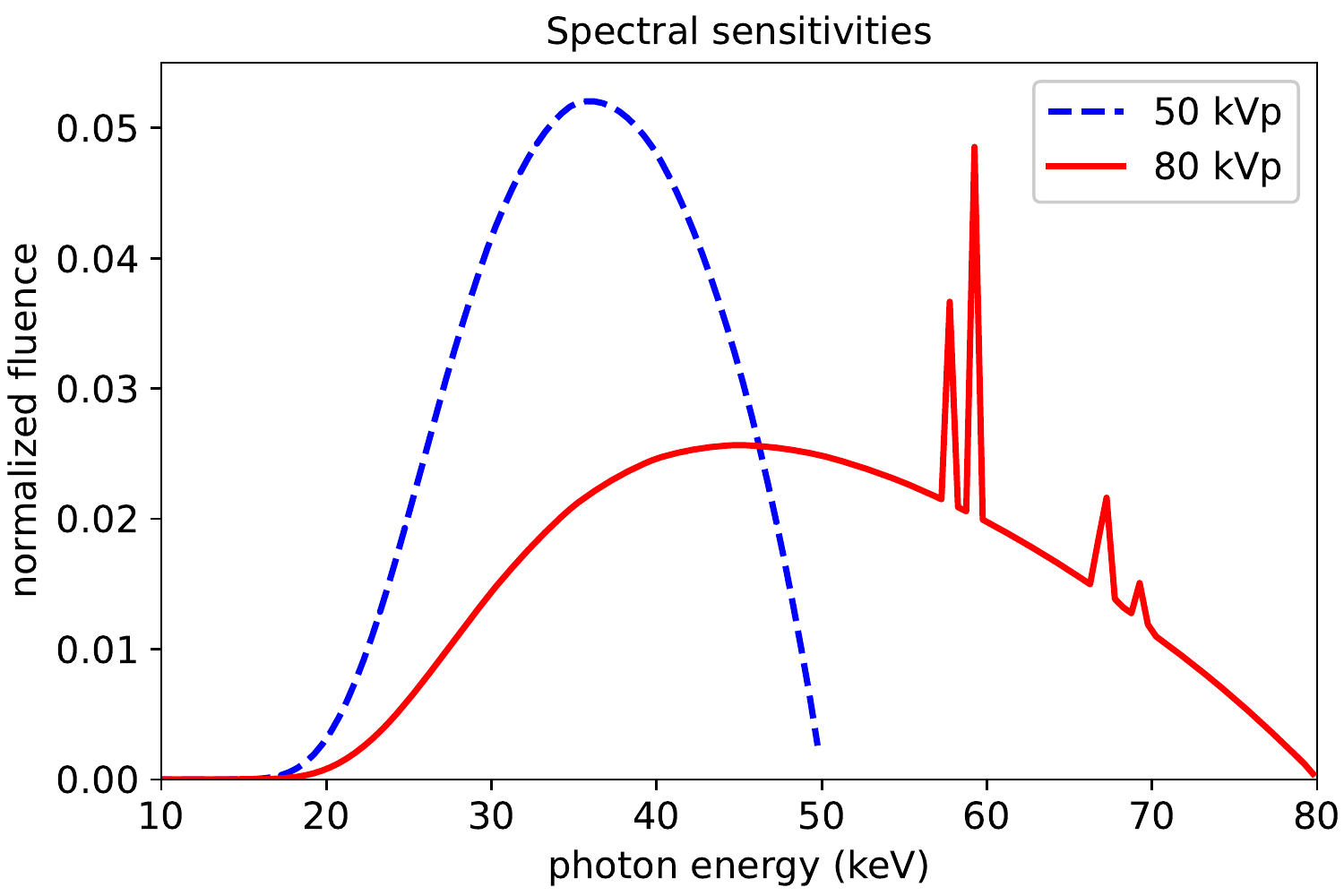}}
\caption{Normalized spectral sensitivities for low and high kVp transmission data. The spectra are generated
modeling a Tungsten anode and 1.6 mm Aluminum filtration. Energy weighting is included in order to model an
energy-integrating detector.
}
\label{fig:spectra}
\end{figure}

\begin{figure}[!t]
\centerline{\includegraphics[width=0.6\columnwidth]{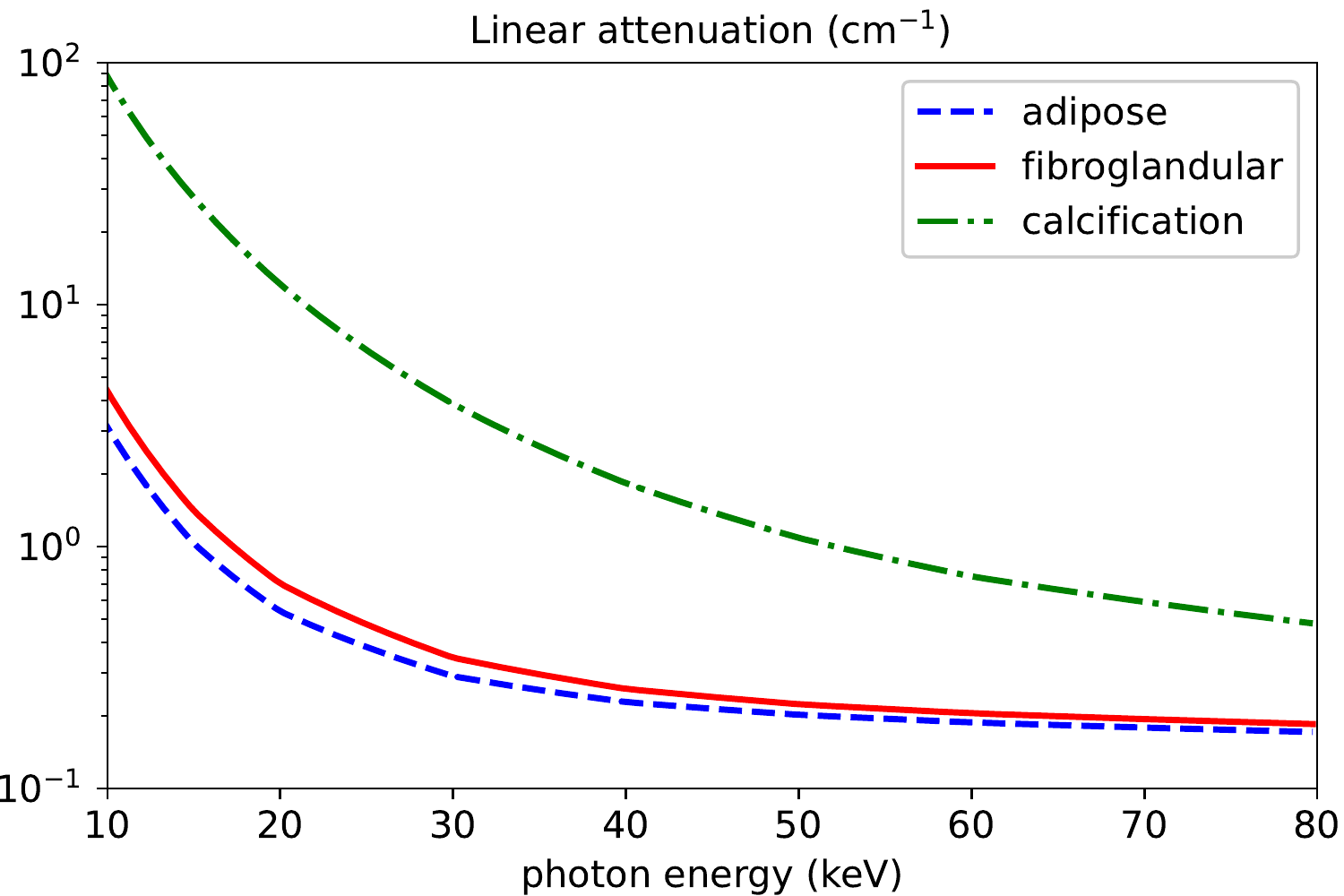}}
\caption{Linear attenuation curves for the three tissues used in the generating the challenge transmission data.
}
\label{fig:lac}
\end{figure}

The breast CT simulation models the transmission data for dual-energy CT as
\begin{equation}
\label{transmissionModel}
I_w = \int s_w(E) \exp \left[-\mu_a(E) P_w x_a - \mu_f(E) P_w x_f - \mu_c(E) P_w x_c \right] dE,
\end{equation}
where $I_w$ is the normalized transmission data for the kVp setting indexed by $w \in \{\text{low},\text{high}\}$;
$s_w(E)$ is the normalized spectral response for kVp setting $w$, i.e.
\begin{equation*}
\int s_w(E) dE = 1;
\end{equation*}
$\mu_a(E)$, $\mu_f(E)$, and $\mu_c(E)$ are the linear attenuation coefficients for adipose, fibroglandular,
and calcification tissues, respectively, at X-ray energy $E$; $P_w$ represents X-ray projection and the
index $w$ is included because the projection matrix is different for high and low kVp settings; and $x_a$, $x_f$, and
$x_c$ are the spatial distributions of adipose, fibroglandular, and calcification tissues, respectively.
By using normalized spectra, the transmission through air is also normalized to 1.
For the dual-energy setup we select the settings to be 50 and 80 kVp for the low and high energy scans, respectively.
The X-ray source spectra are generated using the SpekPy software described in Bujila {\it et al.} \cite{bujila2020validation}.
The $s_w(E)$ functions shown in Fig.~\ref{fig:spectra}
are obtained by multiplying the X-ray source spectra by $E$, assuming an ideal
energy-integrating detector, and normalizing the resulting distribution. The linear attenuation coefficients
shown in Fig.~\ref{fig:lac}
are obtained from the mass attenuation coefficients available from Hubbell and Seltzer \cite{Hubbell:95}
multiplied by the density of the respective tissue type.

For the generation of the tissue sinograms, $P_w x$, we implement a standard line-intersection model for the X-ray
transform of a discrete image.
The images $x$ are pixelized on a 512x512 grid with physical area (18cm)$^2$. The X-ray source follows a circular
trajectory with a radius of 50 cm and centered on the middle of the image. The source-to-detector-center distance
is 100cm, and the detector is modeled as a linear array, which is the exact length needed to capture rays passing
through the largest inscribed circle of the image array. The low and high kVp sinograms are comprised of 256 equally
distributed views over a 360 degree scan and the projections are sampled on a 1024-pixel detector.
The low and high kVp sinograms are differentiated by the fact that their projection view angles
are shifted by a half of the angular sampling interval, thereby simulating a 512-view acquisition where the kVp is switched
for consecutive views.

\begin{figure}[!t]
\centerline{\includegraphics[angle=90,width=0.9\columnwidth]{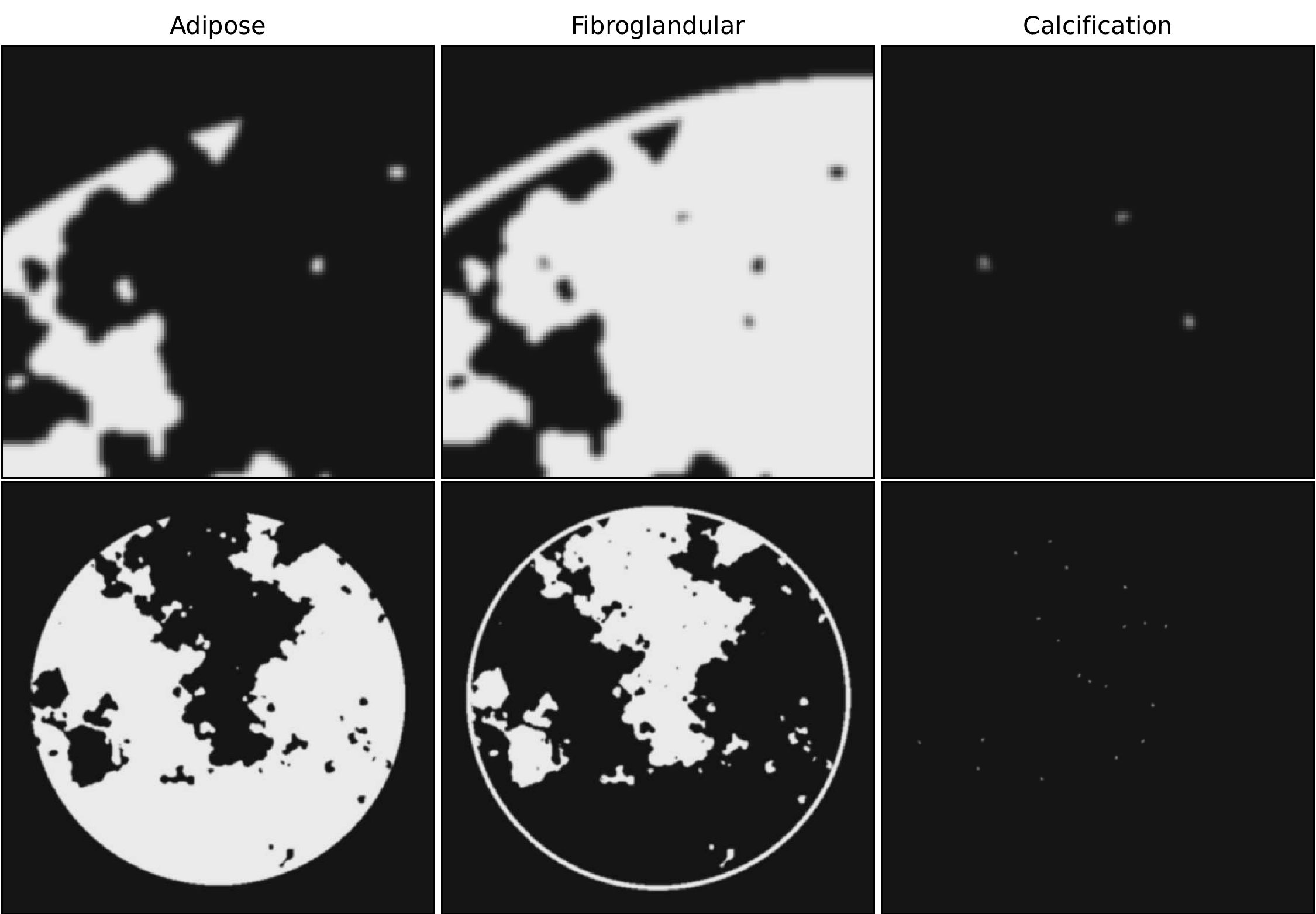}}
\caption{One realization of the breast phantom model. Shown are the three tissue maps with a blow-up of an ROI
in the top row and the complete image in the bottom row. The gray scale range is [-0.1,1.1]. The voxel values, themselves,
range from 0 to 1, indicating the fill fraction of the particular tissue in the pixels.
}
\label{fig:phantoms}
\end{figure}

As with the 2021 DL-sparse-view CT challenge, a stochastic breast model is used to generate multiple
phantoms for training, validating, and testing purposes in DL-spectral CT.
The method for generating the fibroglandular tissue spatial distributions are based on work by Reiser and
Nishikawa. \cite{Reiser10,sidky2020cnns}
The skin line is included with the fibroglandular tissue map and the skin line is taken to be a fixed circular
ring. To create this tissue map, pixels that are within the fibroglandular region or skin line are set to 1
and all other pixels are set to zero. This map is then smoothed with a Gaussian kernel to create a smooth
transition at the edges of the distributions. The adipose tissue map is specified by 
subtracting the fibroglandular map from a uniform disk within the skin line.
A calcification map
with smoothed specks of random shapes are placed only within the fibroglandular regions. This calcification map
is subtracted from the fibroglandular tissue map. All tissue maps range from zero to one, indicating
the volume fraction within each pixel. The sum of all the
tissue maps is exactly equal to one in the region within the circular skin line.
An example realization of the breast phantom tissue map distributions is shown in Fig. \ref{fig:phantoms}.

\begin{figure}[!t]
\centerline{\includegraphics[width=0.6\columnwidth]{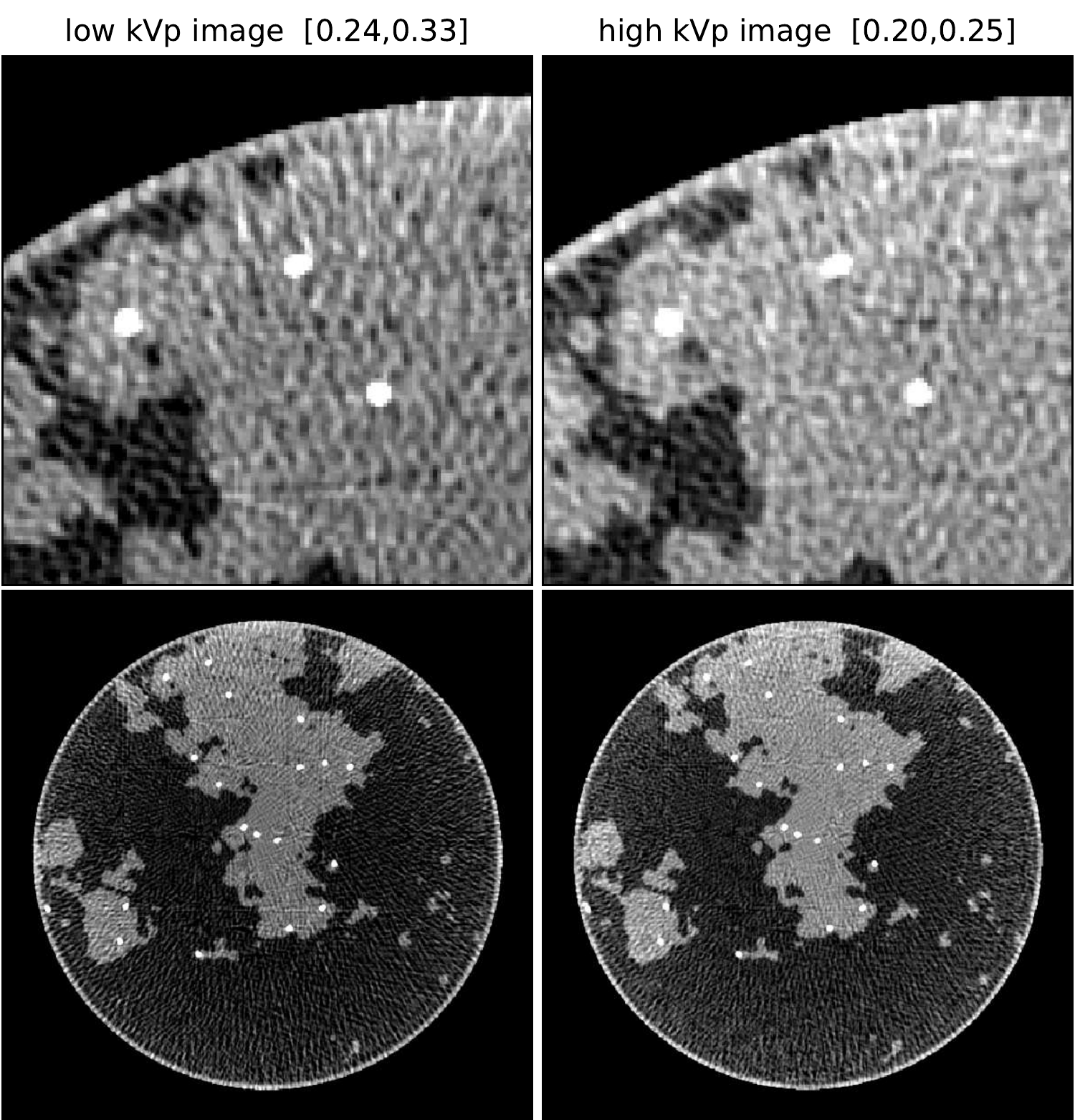}}
\caption{Dual energy images reconstructed by applying FBP to the dual-energy sinograms.
The ROI shown in the top row is the same as the ROI shown in Fig.~\ref{fig:phantoms}. The gray scale
in cm$^{-1}$ is indicated in the column titles. Note the cupping artifact, where the image center is slightly
darker than the outer rim of the test phantom. This cupping, a.k.a. beam-hardening,
is due the fact that the logarithm processing
does not exactly invert the exponentiation in the transmission model when the spectra are broad. The streak
artifacts originate from the discrete angular sampling where only 256 views are taken for each of the kVp
settings.
}
\label{fig:FBPinput}
\end{figure}

The transmission data are generated from the breast phantom tissue maps by Eq. (\ref{transmissionModel});
no additional inconsistency, such as noise, is introduced into the transmission data.
For participants who
want to pursue an image-to-image approach images are generated from $I_\text{low}$ and $I_\text{high}$
by applying FBP to the dual-energy sinograms, which are obtained by
\begin{equation}
\label{desino}
g_w = - \log I_w,
\end{equation}
where $I_w$ is defined in Eq.~(\ref{transmissionModel}).
The resulting low and high dual-energy images
corresponding to the realization shown Fig.~\ref{fig:phantoms} is shown in Fig.~\ref{fig:FBPinput}.
Algorithms developed in the DL spectral CT challenge can use the transmission data, dual-energy sinograms,
or the dual-energy images as inputs to their algorithm.

\begin{figure}[!t]
\centerline{\includegraphics[width=0.6\columnwidth]{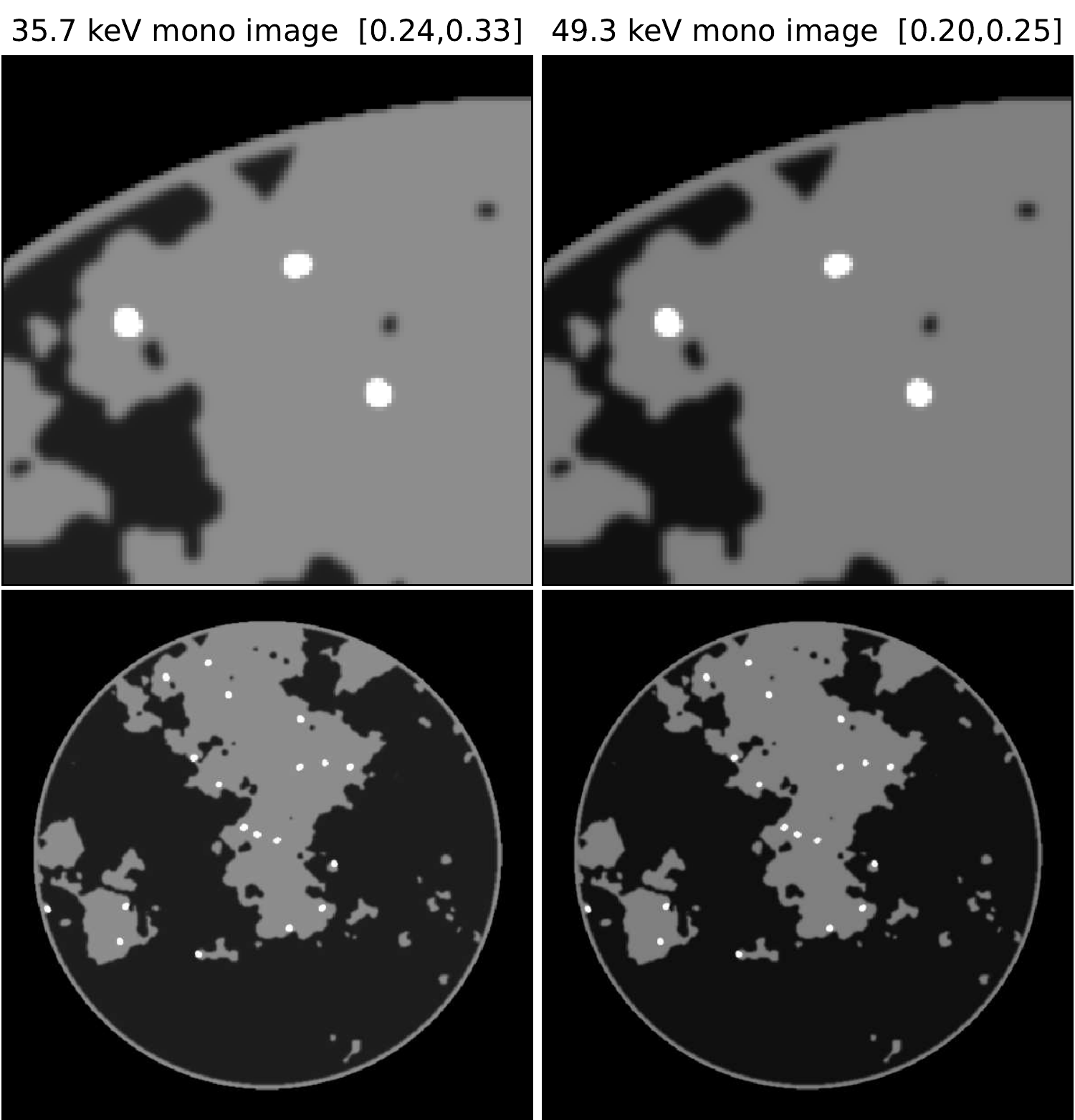}}
\caption{Virtual monochromatic images (VMIs) corresponding to the dual-energy images shown in Fig.~\ref{fig:FBPinput}
using the same gray scales. The left/right columns show the VMIs for 35.7/49.3 keV X-ray photons, which
are the mean energies of the low/high kVp spectra.}
\label{fig:mono}
\end{figure}

An important concept in dual-energy imaging is what is known as the virtual monochromatic image (VMI),
and VMIs play a role in this challenge because many of the participants 
estimate VMIs as an intermediate step on the way to estimating the tissue maps.
The VMI is a hypothetical construct, where the spectral response is a Dirac delta function
\begin{equation*}
s^\text{VMI}_w(E) = \delta(E-E_w).
\end{equation*}
Substituting this spectrum model into Eq.~(\ref{transmissionModel}), the VM sinograms can be found using
Eq.~(\ref{desino})
\begin{equation*}
g^\text{VMI}_w = P_w \; x^\text{VMI}(E_w),
\end{equation*}
which is the projection of the VMI
\begin{equation}
\label{vmi}
x^\text{VMI}(E_w) = \mu_a(E_w) x_a + \mu_f(E_w) x_f + \mu_c(E_w) x_c.
\end{equation}
For illustration, the VMIs, shown in Fig.~\ref{fig:mono}, are formed by performing a
weighted sum of the tissue maps in Fig.~\ref{fig:phantoms}.
The advantage of using the VMIs as an intermediate estimation step
is that going from the dual-energy images in Fig.~\ref{fig:FBPinput}
to the VMIs in Fig.~\ref{fig:mono} amounts to removal of image artifacts for which there many
U-net-based CNNs published in the literature.

\subsection{Challenge logistics and scoring metrics}

The provided training set consisted of 1,000 cases, where each case included the three ground truth tissue maps,
dual-energy transmission data, and dual-energy FBP images. Participants were tasked with developing an algorithm
that can generate the tissue maps from the dual-energy transmission data or FBP images. There was no restriction
on the type of algorithm; algorithms could be iterative, neural-network-based, or a hybrid of the two.
Also provided were python codes for generating transmission data from the tissue maps and the FBP images
from the dual-energy transmission data. In this way, all participants had access to exact knowledge of the
dual-energy CT modeling.

The scoring of the submissions use two RMSE-based metrics.
The RMSE averaged over all 100 test case predictions of the tissue maps is the primary means of 
determining the ranking of submissions
\begin{equation*}
s_1 = \frac{1}{100} \sum_{i=1}^{100} \sqrt{ \frac{ \|t_i-r_i\|^2_2 } {3n} },
\end{equation*}
where $r$ and $t$ are the triplet image sets of reconstructed and ground truth tissue maps, respectively;
$i$ is the test case index; $n=512^2$ is the number
of pixels in a single tissue map image.
In case of a tie-breaking situation and for further characterization of the results,
we also computed a worst-case ROI RMSE
\begin{equation*}
s_2 = \max_{i,c} \sqrt{ \frac{b_c^\top \|(t_i-r_i)\|^2_2 } {3m} },
\end{equation*}
where $b_c$ is the image of an indicator function for a 25x25 pixel ROI centered on coordinates $c$, and $m=625$
is the number of pixels in the test ROI for each tissue map. The location for the test ROI is the same for
each of the tissue map images.
For this competition, it was not clear from the outset that exact recovery was possible, which would be indicated if $s_1=s_2=0$.
As of yet, there is no published image reconstruction algorithm that is capable of solving the posed inverse
problem.

As with the previous 2021 DL sparse-view CT challenge,
the 2022 DL spectral CT challenge\cite{dlspectral} is hosted by the MedICI Platform
(see \url{https://www.medici-challenges.org/}). The MedICI team hosts the challenge
data, which is accessible by sftp protocol, and they also provided the initial
implementation of the challenge on CodaLab
(see \url{https://github.com/codalab/codalab-competitions/wiki/Project_About_CodaLab}).
Using CodaLab, the MedICI team created the challenge and leaderboard format, set up the submission checking, and
created the python plugins that perform the computation of the challenge metrics.
The CodaLab challenge page for DL spectral CT ran the competition in automated fashion
dividing the competition into three phases: training, validation, and testing.
%The authors of this
%challenge report
%were given access to the CodaLab source page in order to edit the text of the DL sparse-view
%CT Challenge website.
%Upon release of the training data, the challenge website ran in a completely
%automated fashion allowing for participant submission of training, validation, and test results
%in the announced time windows of each phase of the challenge.

The training data were released on March 17th to start off the 2022 DL-spectral CT challenge.
On March 31st, 10 new cases were made available, where only the dual-energy FBP images and transmission data
were released. Participants could submit their predicted tissue maps for these 10 cases as many times
as they wished, and a leaderboard marked the progress of the competition by
showing the ranking of the scores for these 10 cases. For this validation phase, participants were
not required to submit their results to the leaderboard.
The final test phase of the competition ran from May 17th to May 31st. For this phase,
100 new cases were generated and the ground truth tissue maps were withheld.
Participants could only submit results for evaluation three times,
and the scores for each test phase submission was automatically registered on the leaderboard.
In this way, all participants were aware of the best scores submitted.
In total, 18 teams submitted test phase results and a total of 39 submissions were made, with 14 submissions
sent on the last day of the competition, and 5 in the last hour. The eventual second place result was submitted
at 10:20 pm on the last day, and the winning result came in at 11:24 pm. With the late lead change and volume
of last day submissions, the drama of the competition was high!

\section{Results}
\label{sec:results}

\begin{figure}[!t]
\centerline{\includegraphics[width=0.5\columnwidth]{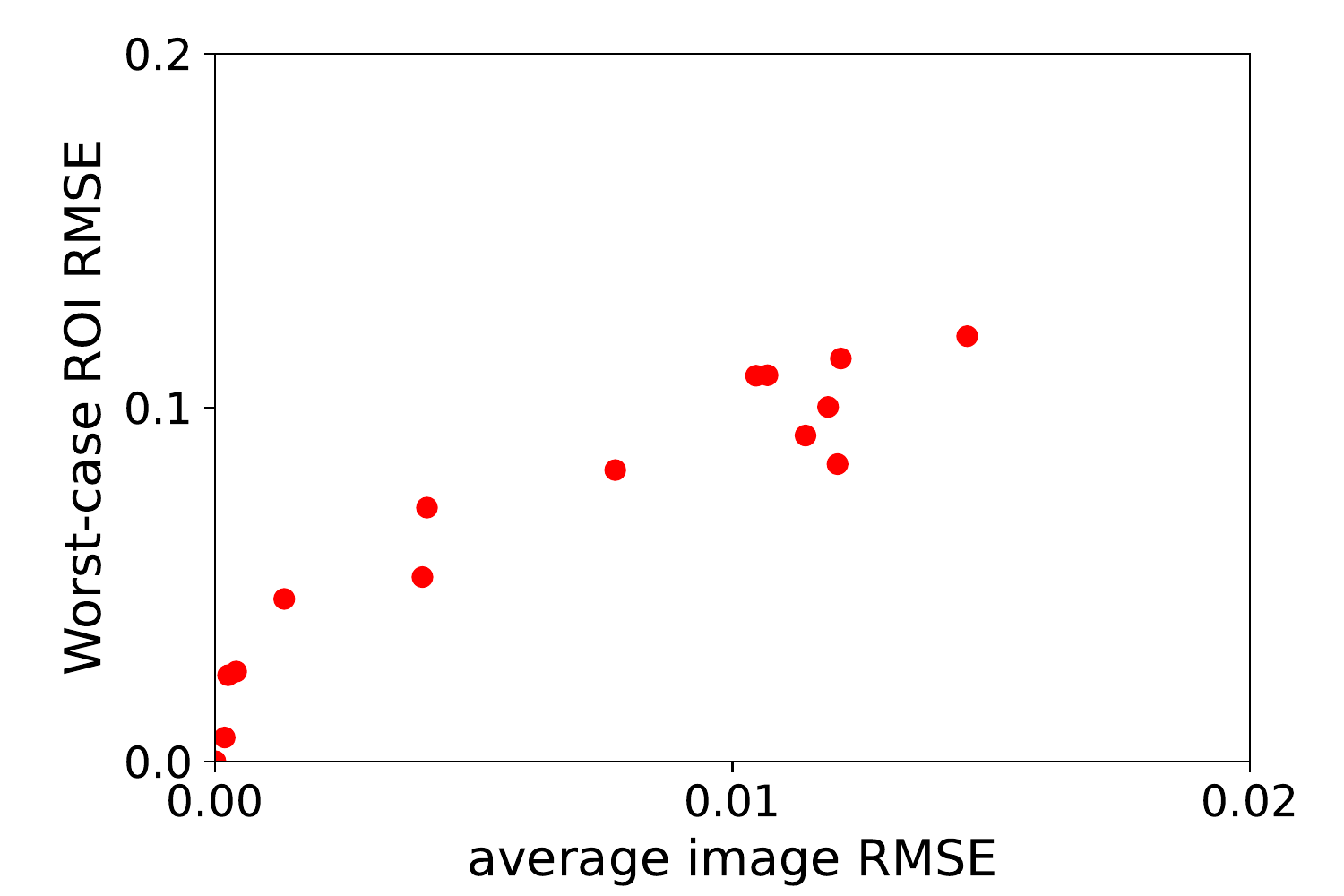}}
\caption{Scatter plot of test phase results with average RMSE $s_1$ on the $x$-axis
and worst-case ROI RMSE $s_2$ on the $y$-axis.
}
\label{fig:testphaseSC}
\end{figure}

\begin{figure}[!t]
\centerline{\includegraphics[width=0.5\columnwidth]{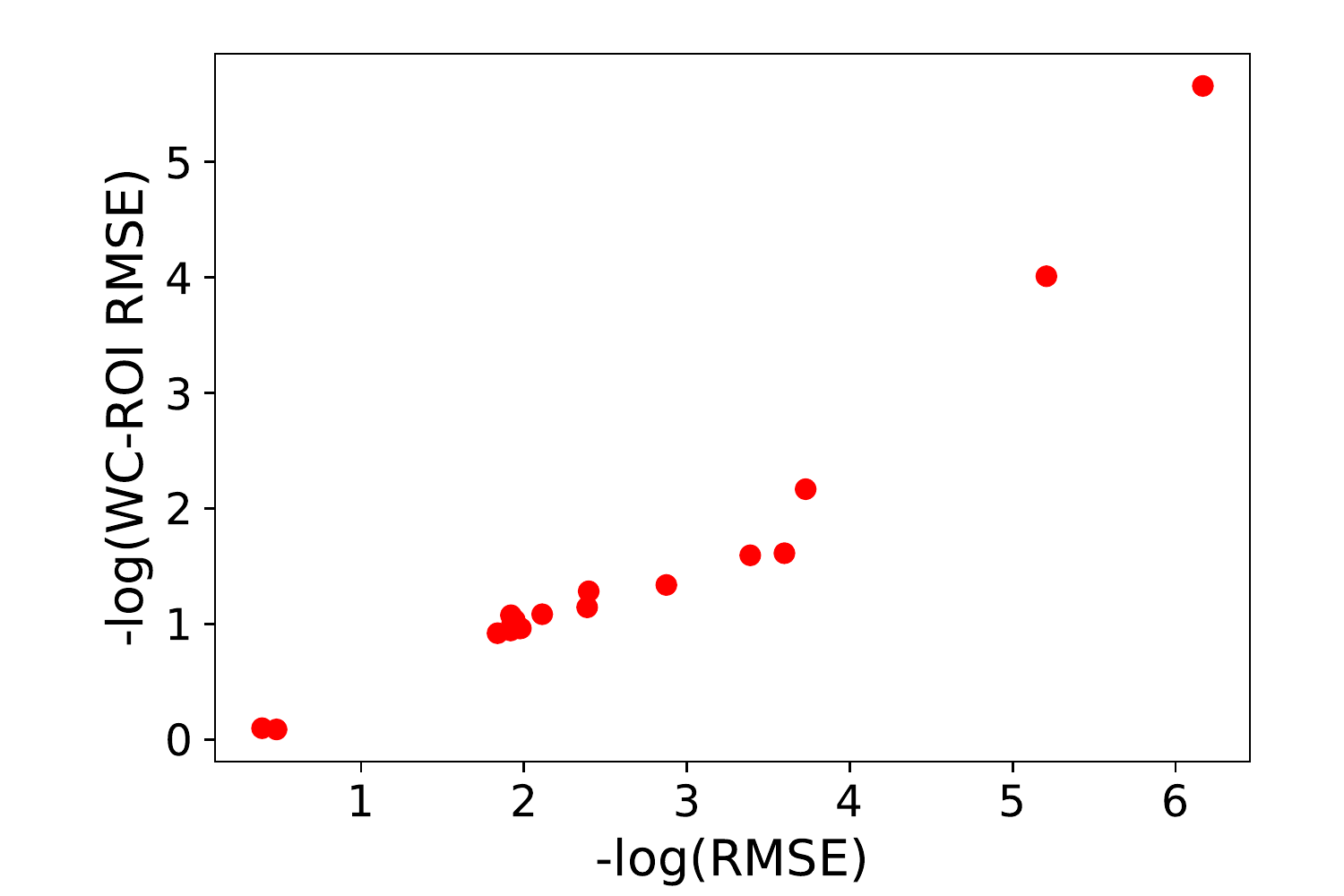}}
\caption{Scatter plot of the same test phase results shown in Fig.~\ref{fig:testphaseSC}
except that the RMSE results are plotted after processing with the negative base-10 logarithm;
i.e., the $x-$ and $y$-axes are  $-\log_{10}s_1$ and $-\log_{10}s_2$, respectively.
In this way, the top performers, which appear in the upper right corner of the plot, are more
clearly differentiated.
}
\label{fig:testphaselogSC}
\end{figure}

Scatter plots of the final test phase results appear in Figs.~\ref{fig:testphaseSC} and \ref{fig:testphaselogSC}.
For the leading results the worst-case ROI RMSE tracks with the average RMSE, which is the metric used
for ranking the submitted tissue map predictions. Generally speaking the performance level for the
challenge participants was excellent; the natural scale of this reconstruction problem is zero to one,
which are the extreme ranges of the tissue filling fraction for the image pixels. The obtained RMSE
values are substantially below one, and the leading results approach numerical exactness at the
level of four-byte, i.e. single precision, floating point arithmetic. The scatter plot in Fig.~\ref{fig:testphaseSC}
reveals a healthy spread of scores, and the worst-case ROI RMSE is generally an order of magnitude
larger than the corresponding mean RMSE as might be expected. Due to the large dynamic range in scores,
the two leading results appear at the origin of this plot. To focus on the leading results, in Fig.~\ref{fig:testphaselogSC}
we also show the same scoring data using the negative base-10 logarithm of the RMSE values.
The degree of the prediction accuracy of the top two performers becomes clear in this plot.
The spread in mean RMSE values allowed unambiguous ranking, nevertheless we also provide the worst-case ROI RMSE
values to demonstrate the high level of accuracy in the tissue map predictions.
The top ten teams are listed in Table \ref{tab:winners}, where the numerical scores are also shown.
A full listing of all results are available on the DL spectral CT challenge website.

\begin{table}
\begin{center}
\begin{tabular}{c| c c} 
 \hline
 Username/team &  $s_1$ & $s_2$   \\ [0.5ex] 
 \hline\hline
 GenwaiMa/GM\_CNU           & 6.80$\times 10^{-7}$  & 2.20$\times 10^{-6}$   \\ 
 \hline
 huxiaoyu090/iTORCH         & 6.21$\times 10^{-6}$  & 9.77$\times 10^{-5}$   \\
 \hline
 kimhs369/MIR               & 1.87$\times 10^{-4}$  & 6.81$\times 10^{-3}$   \\
 \hline
 WashUDEAM                  & 2.52$\times 10^{-4}$  & 2.44$\times 10^{-2}$   \\
 \hline
 dhlee91                    & 4.08$\times 10^{-4}$  & 2.54$\times 10^{-2}$   \\
 \hline
 jaspernijkamp/DCPT+Navrit  & 1.34$\times 10^{-3}$  & 4.60$\times 10^{-2}$   \\
 \hline
 Duke\_QIAL                 & 4.01$\times 10^{-3}$  & 5.21$\times 10^{-2}$   \\
 \hline
 leekunpeng/BME\_NUC        & 4.10$\times 10^{-3}$  & 7.18$\times 10^{-2}$   \\
 \hline
 flutexu                    & 7.73$\times 10^{-3}$  & 8.24$\times 10^{-2}$   \\
 \hline
 Z-VCT                      & 1.04$\times 10^{-2}$  & 1.09$\times 10^{-1}$   \\ [1ex] 
 \hline
\end{tabular}
\end{center}
\caption{Top ten participating teams and their scores. First place is at the top of the list.
For viewing all results, please visit the DL spectral CT challenge website:
\url{https://dl-sparse-view-ct-challenge.eastus.cloudapp.azure.com/competitions/3}
\label{tab:winners}}
\end{table}

One of the more surprising results of the DL spectral CT challenge is the variety of approaches taken
by the challenge participants, and the fact that highly accurate results were obtained with vastly different
approaches. The original plan for this challenge report was to highlight the top five
algorithms, but the number of high-quality submissions at the test phase forced us to reconsider this plan.
We highlight the top ten performers, and we also acknowledge that drawing the line at the top ten
is rather arbitrary.

The algorithms developed by the top ten participating teams are briefly summarized, and the individual teams
are encouraged to publish full presentation of their algorithms. The team membership and institutions are listed.
The e-mail address for the contact member is also provided. With each algorithm description, images
of the tissue map worst-case ROI are shown. The worst-case ROI images are not shown for the second and first place
teams, since there is no visually discernible error in their tissue map predictions.

\subsubsection*{Tenth place}

\begin{figure}[!t]
\centerline{~~~~~Adipose~~~~~~~~~Fibroglandular~~~~~~Calcification}
\centerline{\includegraphics[width=0.6\columnwidth]{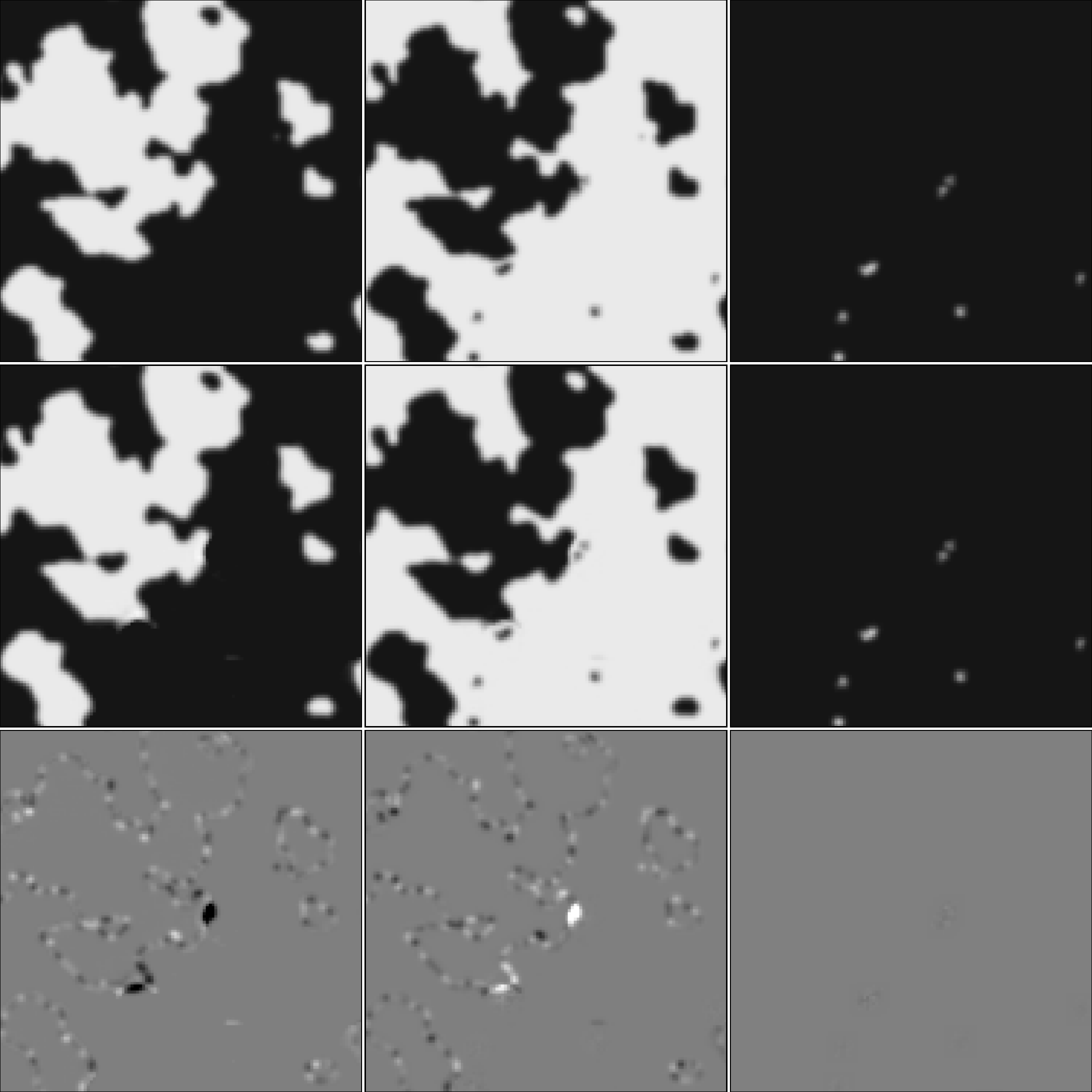}}
\caption{Worst-case ROI for the tenth place team. The worst-case ROI is a 25x25 pixel region
with the maximum discrepancy, over 100 test images, between the true and predicted tissue maps.
For better image context, the shown ROI is expanded 125x125 pixels centered on the worst-case ROI.
The top and middle rows show the ground truth and predicted tissue maps, respectively, in
a gray scale range of [-0.1,1.1]. The bottom row shows the difference in a gray scale window [-0.5,0.5].
}
\label{fig:Z-VCT-ROIs}
\end{figure}

Team Z-VCT uses a combined CNN-based and iterative approach.
Their algorithm consists of three major components. The first step seeks to
obtain dual-energy images with reduced artifacts from the dual-energy sinograms.
For this stage, the SART algorithm is implemented using non-local means (NLM)
regularization, thereby mitigating the streak artifacts seen in the provided
FBP-based dual-energy images. The second stage is composed of two neural networks,
the first of which converts the two dual-energy images into an estimate
of the three tissue maps. The second network refines this estimate using
a standard U-Net. The final, third stage exploits the provided dual-energy physics
information with a nonlinear algebraic iterative algorithm, where the tissue
maps estimated from the second stage are used to initialize the iteration,
in this way the final tissue map estimates are made consistent with the
provided dual-energy transmission data.
Team Z-VCT members are Yizhong Wang (wyzwys0101@163.com) and Ailong Cai; both members are
from the Information Engineering University in Zhengzhou, China.

The worst-case ROI for this algorithm is shown
in Fig.~\ref{fig:Z-VCT-ROIs}. The tissue map predictions for this algorithm are visually
very accurate; only with guidance from the difference image is it possible to observe small
discrepancies in the ground truth and predicted ROI images.
The difference images reveal that the largest errors are located at the tissue
borders, mainly for the adipose and fibroglandular maps,
and that the errors in these two maps are anticorrelated. The form of these errors likely
result from the fact that the tissue maps have fractional values at the borders
and the fact that the adipose and fibroglandular linear attenuation coefficients
have quite similar energy dependences as seen in Fig.~\ref{fig:lac}.

\subsubsection*{Ninth place}
\begin{figure}[!t]
\centerline{~~~~~Adipose~~~~~~~~~Fibroglandular~~~~~~Calcification}
\centerline{\includegraphics[width=0.6\columnwidth]{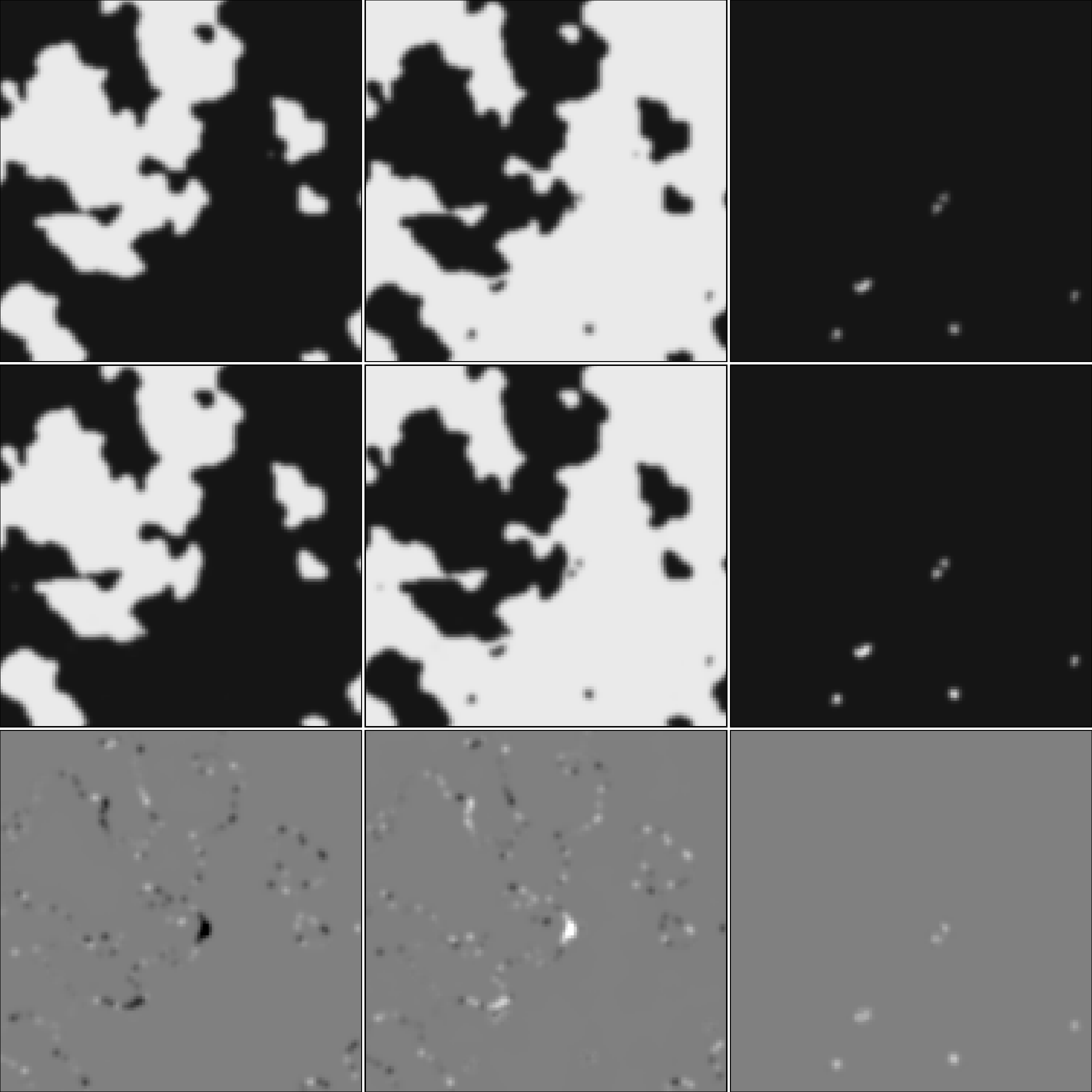}}
\caption{Worst-case ROI for the ninth place team. See Fig.~\ref{fig:Z-VCT-ROIs} for description of ROI image specification.
}
\label{fig:flutexu-ROIs}
\end{figure}

The algorithm developed by team flutexu involves a two-stage
processing chain where the dual-energy sinograms are
reconstructed into dual-energy images, which are subsequently fed into
a neural network to provide estimates of the tissue maps.
The first stage uses an iterative update inspired by the winning team of
the 2021 DL sparse-view CT challenge \cite{genzel2022near}, where the images
are updated based on FBP applied to the difference between the dual-energy sinograms
and their estimate derived from the current iterate of the tissue maps.
The dual-energy FBP images, provided in the challenge, are used to initialize this iteration.
The resulting images of the iterative update are fed into a U-Net CNN to provide
the estimate of the tissue maps. The free parameters of both stages are trained
simultaneously using the provided training cases. This algorithm does exploit
the knowledge of the system geometry in the first stage, but the provided
dual-energy physics model is not explicitly used.
Team flutexu members are Di Xu (DiXu@mednet.ucla.edu) and Ke Sheng; both members are
from the Department of Radiation Oncology at the University of California Los Angeles, Los Angeles,
California, USA.

The worst-case ROI for this algorithm is shown
in Fig.~\ref{fig:flutexu-ROIs}. Interestingly, the same ROI as that of Fig.~\ref{fig:Z-VCT-ROIs}
yielded the worst-case ROI for this algorithm. In fact, this same ROI proved to be the most
problematic for many of the other top ten performing algorithms. Similar observations can
be made about the error distributions of this algorithm as with the previous one. One notable
difference is that the error in the calcification map is visible and, therefore, of similar
amplitude as the error in the other two tissue maps. For the previous result, the error
was mainly occurring in the adipose and fibroglandular tissue maps.

\subsubsection*{Eighth place}
\begin{figure}[!t]
\centerline{~~~~~Adipose~~~~~~~~~Fibroglandular~~~~~~Calcification}
\centerline{\includegraphics[width=0.6\columnwidth]{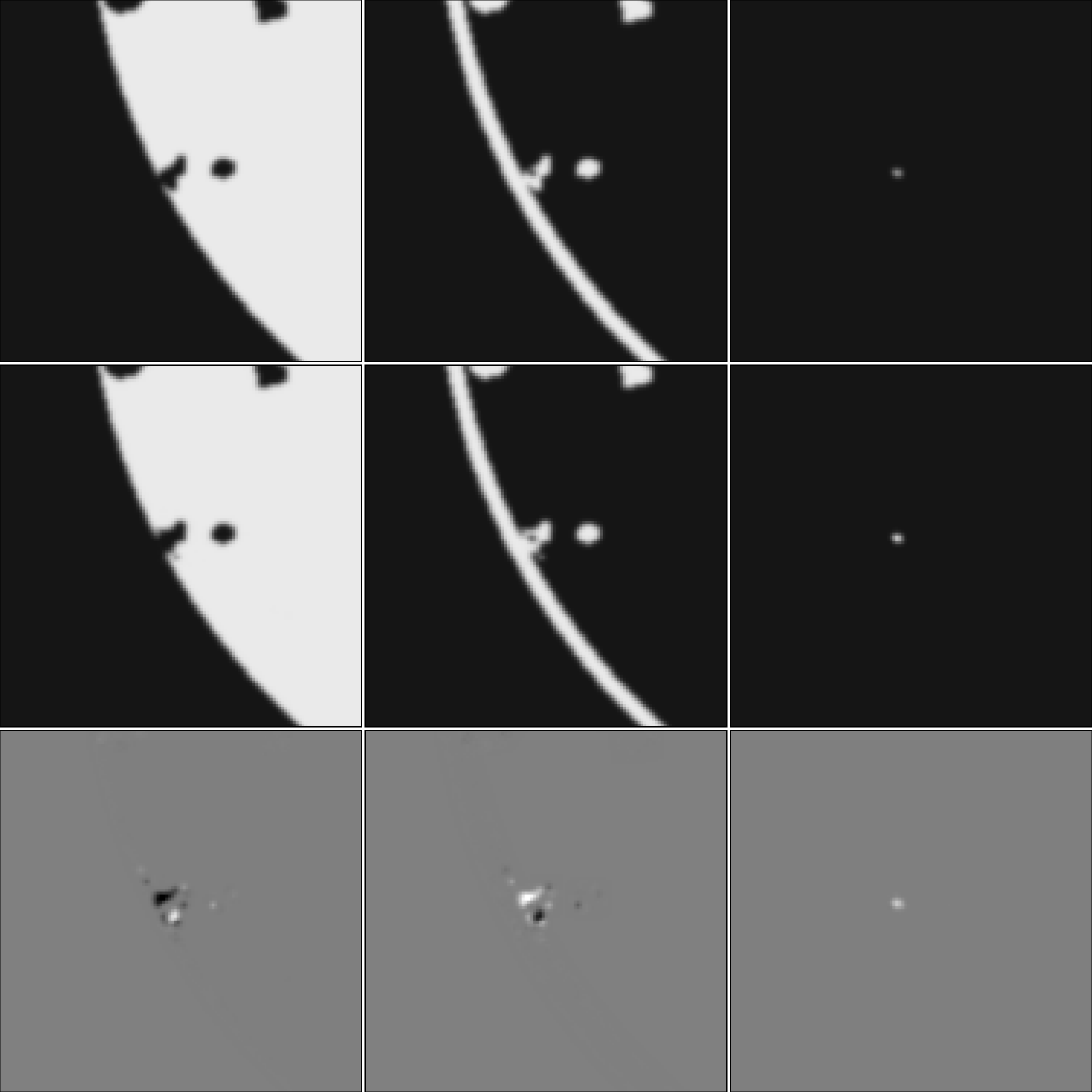}}
\caption{Worst-case ROI for the eighth place team. See Fig.~\ref{fig:Z-VCT-ROIs} for description of ROI image specification.
}
\label{fig:leekunpeng-ROIs}
\end{figure}

Team BME\_NUC developed an algorithm, which they call the
intelligent dual-domain network (iDD-Net). The iDD-Net algorithm
has two CNNs, the first of which takes as input the dual-energy sinograms
and it estimates a fully-sampled sinogram
for each of the tissue maps. In this way, the transformation
from dual-energy data to tissue maps is happening in the sinogram
domain -- a unique feature among the top ten performing algorithms.
The tissue sinograms are back-projected into the image domain
then fed into another CNN to obtain the final tissue map estimates.
The complete iDD-Net is trained end-to-end. The only physical parameters
of the dual-energy set-up used in training iDD-Net are the geometric
parameters needed for the back-projection.
Team BME\_NUC members are Kunpeng Li (leekunpeng@hotmail.com), Pengcheng Zhang, Yi Liu, and Zhiguo Gui; all
members are from the
Shanxi Provincial Key Laboratory for Biomedical Imaging and Big Data, North University of China, Taiyuan, China.

The worst-case ROI for this algorithm is shown
in Fig.~\ref{fig:leekunpeng-ROIs}. For this algorithm, an ROI on the edge of the test phantom yields
the worst-case ROI. The error for this ROI is very localized, as opposed to be distributed along the borders
of the adipose and fibroglandular tissue maps. Anti-correlation is still observed in the two soft-tissue maps, and
an error of comparable amplitude is seen in the calcification map.

\subsubsection*{Seventh place}
\begin{figure}[!t]
\centerline{~~~~~Adipose~~~~~~~~~Fibroglandular~~~~~~Calcification}
\centerline{\includegraphics[width=0.6\columnwidth]{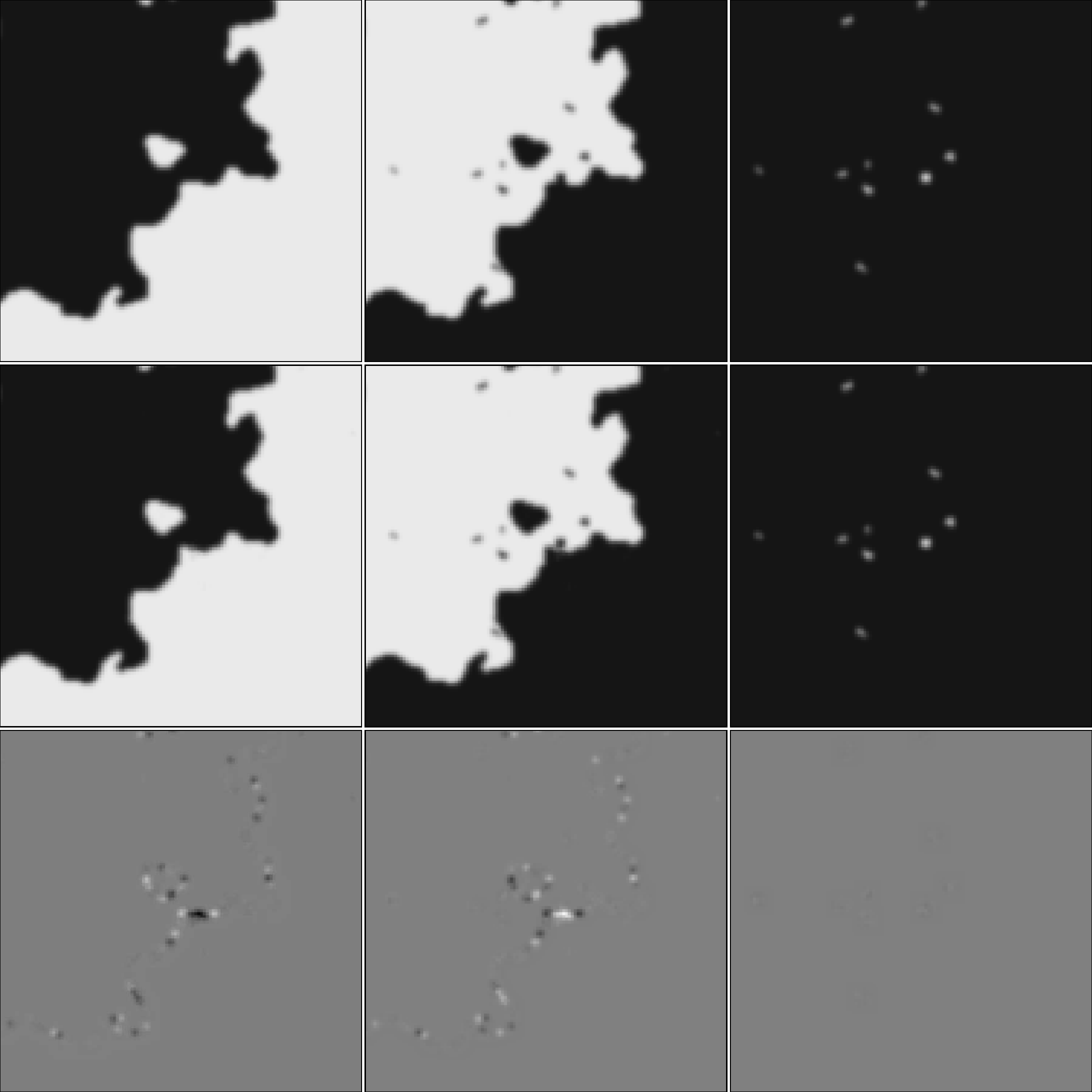}}
\caption{Worst-case ROI for the seventh place team. See Fig.~\ref{fig:Z-VCT-ROIs} for description of ROI image specification.
}
\label{fig:Duke_QIAL-ROIs}
\end{figure}

Team Duke\_QIAL developed a four-stage algorithm involving (1) estimation
of beam-hardening corrected low and high kVp images, (2) a U-net
for estimating tissue maps from the dual-energy images, (3) refinement
by iterative image reconstruction, and (4) a second pass with a U-net.
The first-stage beam hardening correction uses the provided spectral response data
to fit a second order polynomial that relates the dual-energy sinograms
to sinograms consistent with the projection
of VMIs at effective energies for both low and high kVp settings.
These monochromatic sinograms are processed with algebraic reconstruction
to reduce the under-sampling artifacts. In the second-stage a U-net is trained
to convert the two VMIs to the three
tissue maps; the tissue sum constraint and object support are also enforced at
this stage. The third stage runs a single iteration of this team's own
spectral CT image reconstruction algorithm \cite{clark2019photon} to update
the VMIs and to obtain their corresponding residuals.
The final stage involves processing with a U-net similar to the one trained
in the second stage; the difference is that the previous U-net takes
two VMIs as inputs while the final U-net
takes four inputs, VMIs
together with their residuals. For both U-nets, thresholding is used to
set values within 0.5\% of 0 or 1 to exactly 0 or 1.
Team Duke\_QIAL members are Rohan Nadkarni, Darin Clark, and Cristian Badea (Cristian.Badea@duke.edu), and
all members are from the Quantitative Imaging and Analysis Lab, Department of Radiology, Duke University Medical Center, Durham,
North Carolina, USA

The worst-case ROI for this algorithm is shown
in Fig.~\ref{fig:Duke_QIAL-ROIs}. The shown ROI is unique among all of the worst-case ROIs shown for the top
ten performing algorithms. The distribution of tissue map errors is concentrated in the middle of the soft-tissue
ROI images, but there is also some visible discrepancy along the edges of the soft-tissue maps. Again, the soft-tissue
map error is anticorrelated.

\subsubsection*{Sixth place}
\begin{figure}[!t]
\centerline{~~~~~Adipose~~~~~~~~~Fibroglandular~~~~~~Calcification}
\centerline{\includegraphics[width=0.6\columnwidth]{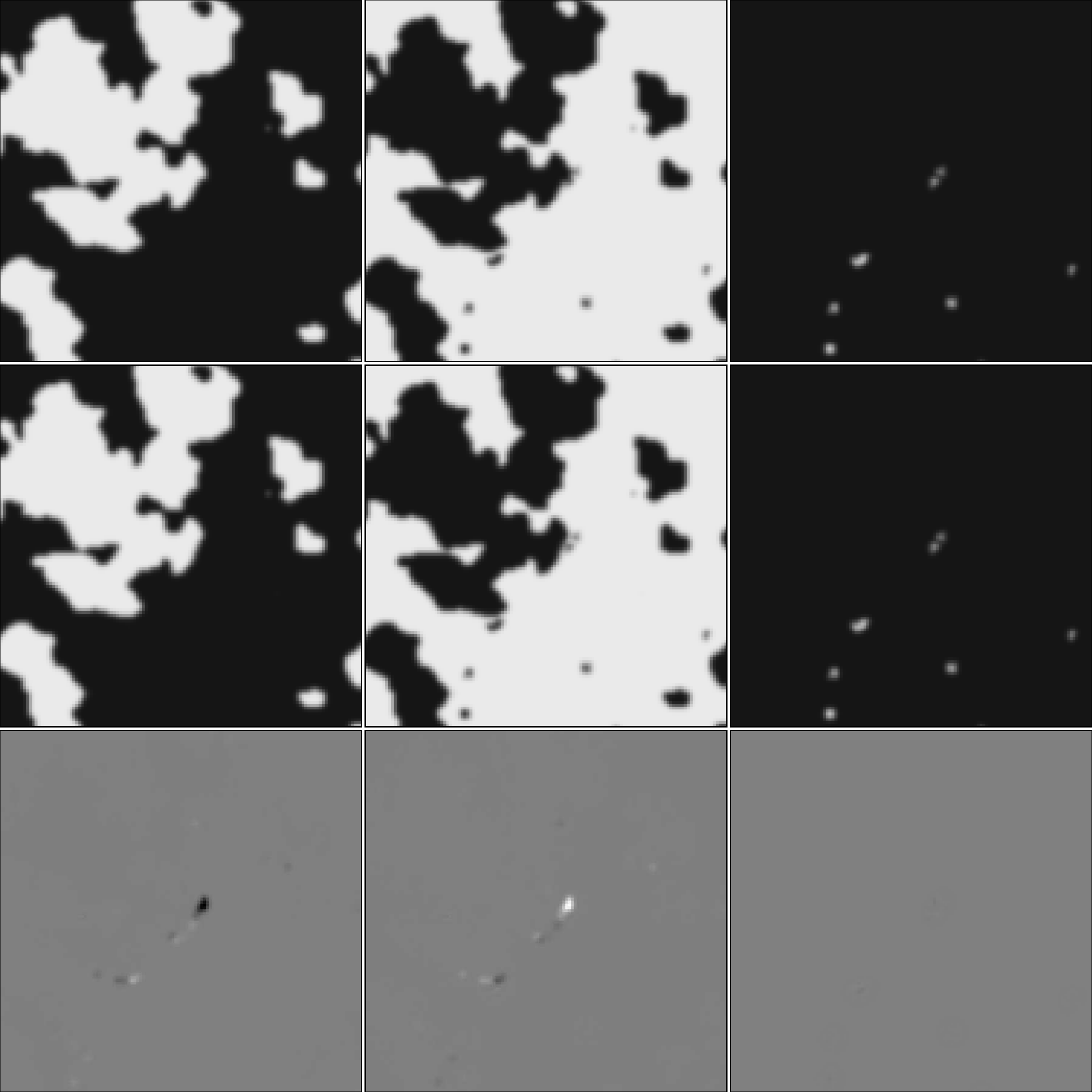}}
\caption{Worst-case ROI for the sixth place team. See Fig.~\ref{fig:Z-VCT-ROIs} for description of ROI image specification.
}
\label{fig:jaspernijkamp-ROIs}
\end{figure}
Team DCPT+Navrit developed a U-net that takes three inputs from which the three
tissue maps are predicted. The inputs are the low kVp image, the high kVp image, and their difference.
The dual-energy images used for the U-net are generated from this team's own
FBP implementation where they implemented a Hann filter instead of the ramp filter used
for generating the dual-energy images in the challenge data set. The U-net is pretrained
using ideal, artifact-free VMIs, which are obtained by fitting the FBP-generated
dual-energy images with linear combinations of the tissue maps in the training set.
After training the U-net on these ideal dual-energy image inputs, transfer learning is used
to subsequently adjust the U-net weights by training on the FBP-generated dual-energy image inputs.
This algorithm had the best performance among all algorithms that did not explicitly use
the given physical spectral model, in Eq.~(\ref{transmissionModel}),
that relates the tissue maps to the dual-energy transmission data.
Team DCPT+Navrit members are Zixiang Wei, Imaiyan Chitra Ragupathy, Jintao Ren, Navrit Bal,
Kristoffer Moos, Mathis Ersted Rasmussen, and Jasper Nijkamp (jaspernijkamp@clin.au.dk).
All members, except Navrit Bal, are from the
Department of Clinical Medicine Aarhus University and
Danish Center for Particle Therapy, Aarhus University Hospital, Aarhus, Denmark.
Navrit Bal is from the Department of Detector Research and Development, National Institute for Subatomic Physics - NIKHEF,
Amsterdam, the Netherlands.

The worst-case ROI for this algorithm is shown
in Fig.~\ref{fig:jaspernijkamp-ROIs}. This same ROI was the worst confounder for the tenth and ninth place finishers;
the error distribution is confined to the soft-tissues and spatially it is quite localized. Anti-correlation in the
soft tissues is observed.

\subsubsection*{Fifth place}
\begin{figure}[!t]
\centerline{~~~~~Adipose~~~~~~~~~Fibroglandular~~~~~~Calcification}
\centerline{\includegraphics[width=0.6\columnwidth]{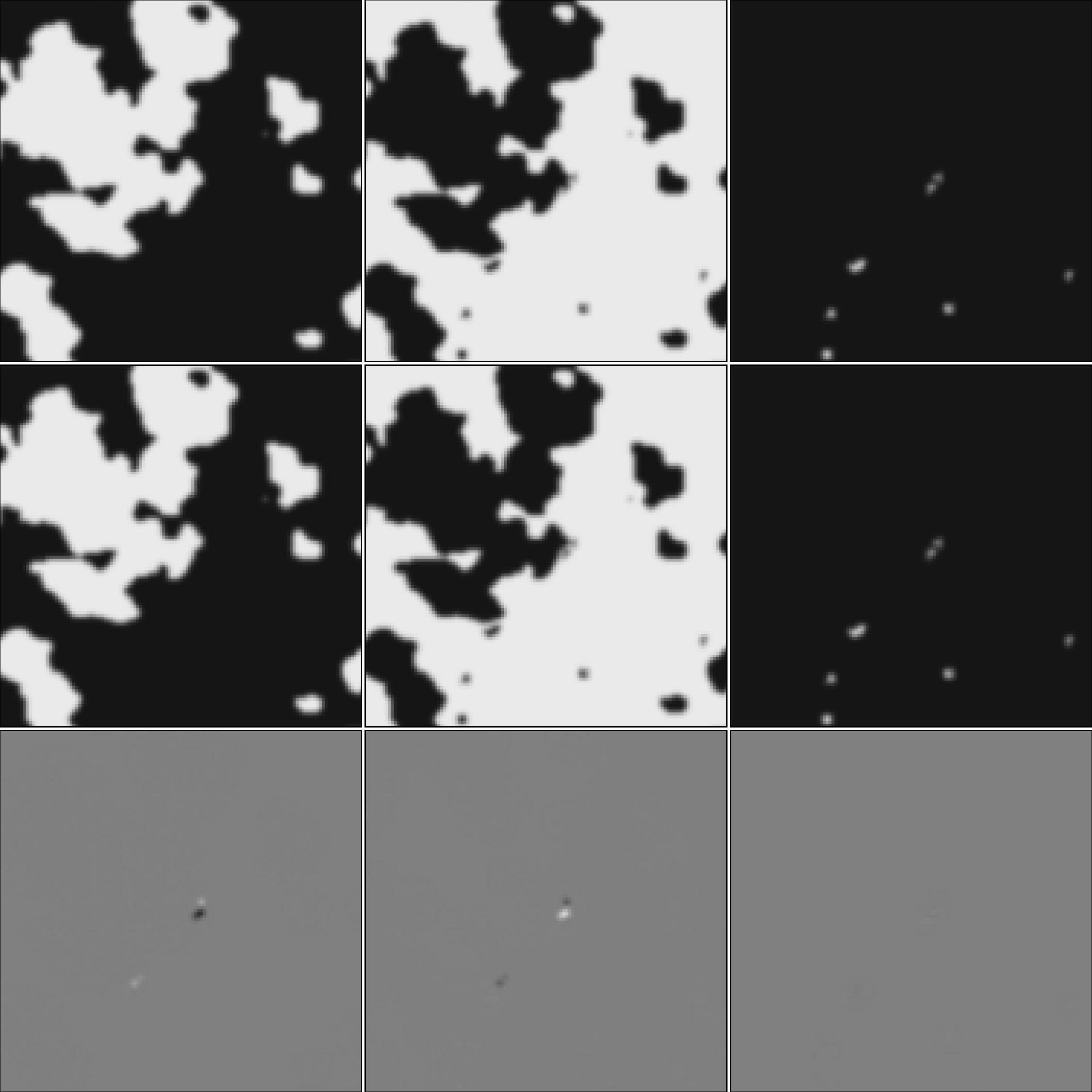}}
\caption{Worst-case ROI for the fifth place team. See Fig.~\ref{fig:Z-VCT-ROIs} for description of ROI image specification.
}
\label{fig:dhlee91-ROIs}
\end{figure}

Team dhlee91 developed an unrolled iterative image reconstruction approach \cite{ongie2020deep},
where subnetworks having the same structure are repeatly used for a fixed number of iterations and
both data consistency steps and regularization steps are replaced with CNNs.
The particular iterative algorithm adopted for the network structure was the primal-dual hybrid gradient
algorithm \cite{chambolle2011first} and regularization was performed in both the sinogram-domain and image-domain.
The algorithm takes the dual-energy sinograms as input and solves for the three tissue maps.
Team dhlee91 had only one member, Donghyeon Lee (dlee258@jhmi.edu) from the
Department of Radiology and Radiological Science, Johns Hopkins University School of Medicine, Baltimore, Maryland, USA.

The worst-case ROI for this algorithm is shown
in Fig.~\ref{fig:dhlee91-ROIs}. The error distribution is similar to that of the sixth place finisher, shown
in Fig.~\ref{fig:jaspernijkamp-ROIs}, except that the visible error is confined to fewer pixels.

\subsubsection*{Fourth place}
\begin{figure}[!t]
\centerline{~~~~~Adipose~~~~~~~~~Fibroglandular~~~~~~Calcification}
\centerline{\includegraphics[width=0.6\columnwidth]{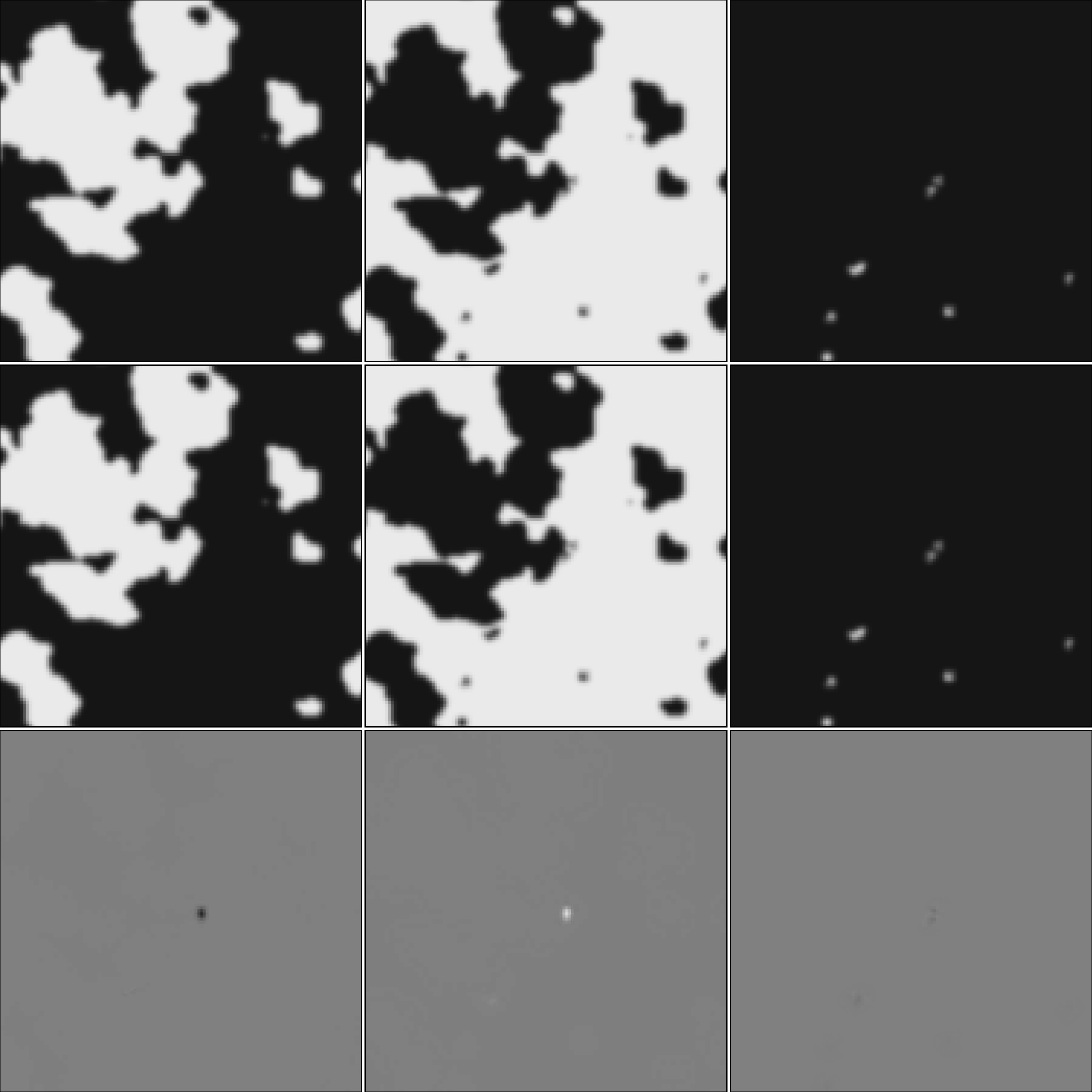}}
\caption{Worst-case ROI for the fourth place team. See Fig.~\ref{fig:Z-VCT-ROIs} for description of ROI image specification.
}
\label{fig:WashUDEAM-ROIs}
\end{figure}

Team WashUDEAM developed a two-stage image reconstruction algorithm. The first
stage is a one-step iterative algorithm that they call dual-energy
alternating minimization (DEAM), and which is based on a published 
algorithms from the same group \cite{o2004alternating,o2007alternating}.
In the implementation of DEAM, two of the three tissues are selected as basis materials,
and the basis maps are solved directly from the dual-energy transmission data
using the given model in Eq.~(\ref{transmissionModel}). In a second stage, a U-net is trained to
estimate the three tissue maps from six inputs derived from DEAM.
The six inputs are obtained from DEAM by selecting all combinations of two tissues as
basis materials, resulting in three possibilities. Each of the three runs of DEAM yields two basis
maps for a total six basis maps that are fed into the U-net. In the U-net implementation
the estimated tissue maps are constrained to values in the interval [0,1].
Team WashUDEAM members are Tao Ge (getao@wustl.edu), Maria Medrano, Joseph A. O'Sullivan, and
all members are from the Department of Electrical and Systems Engineering,
Washington University in St. Louis, St. Louis, USA.

The worst-case ROI for this algorithm is shown
in Fig.~\ref{fig:WashUDEAM-ROIs}. The error distribution is similar to that of the fifth and sixth place finisher, shown
in Figs.~\ref{fig:jaspernijkamp-ROIs} and \ref{fig:dhlee91-ROIs}, except that the visible error is confined to an
even smaller grouping of pixels.

\subsubsection*{Third place}
\begin{figure}[!t]
\centerline{~~~~~Adipose~~~~~~~~~Fibroglandular~~~~~~Calcification}
\centerline{\includegraphics[width=0.6\columnwidth]{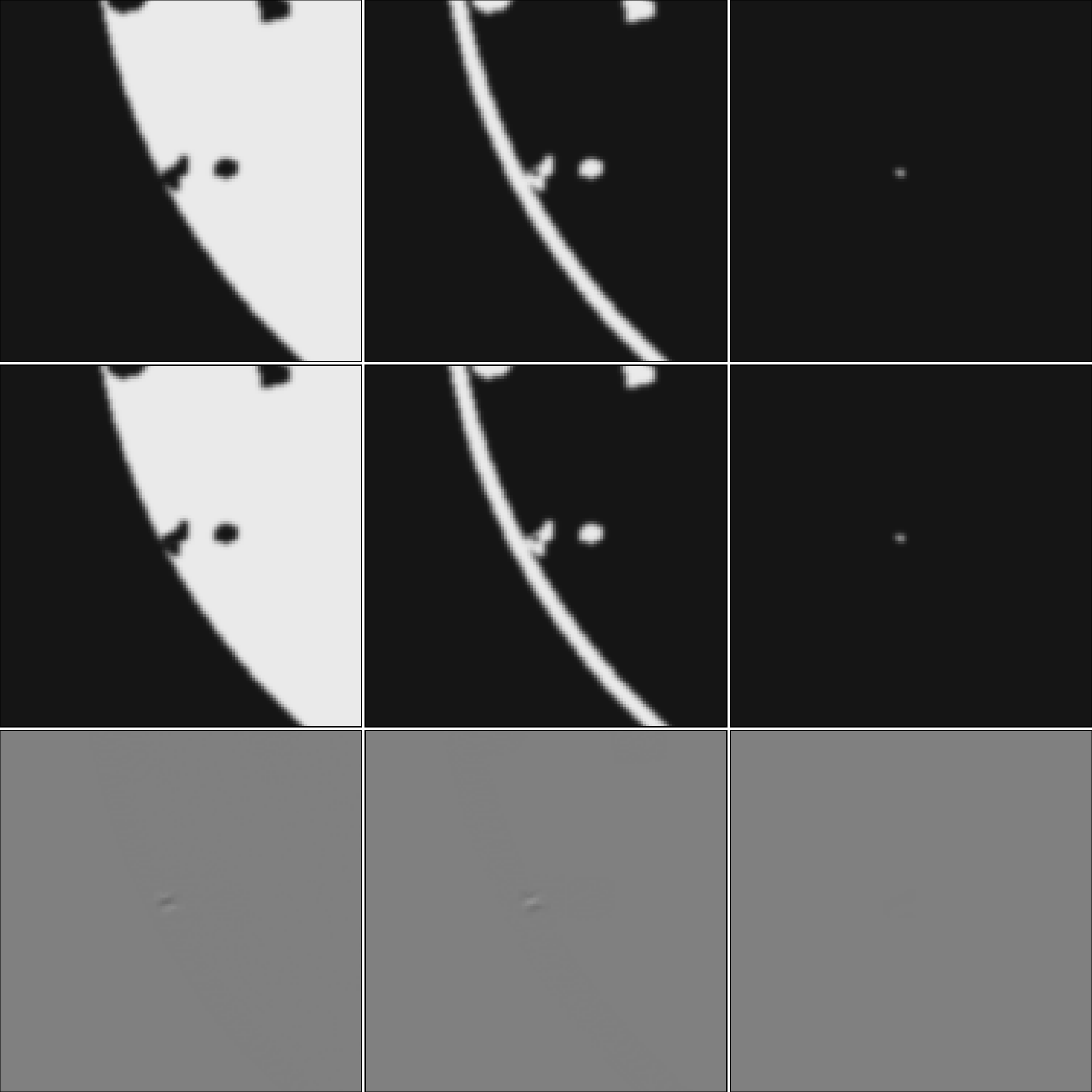}}
\caption{Worst-case ROI for the third place team. See Fig.~\ref{fig:Z-VCT-ROIs} for description of ROI image specification.
}
\label{fig:kimhs369-ROIs}
\end{figure}

The algorithm developed by team MIR employs four U-nets and enforces data consistency
through a simultaneous algebraic reconstruction technique (SART) iteration.
The first network estimates the sum of tissue maps from the dual-energy FBP images.
For the DL-spectral CT challenge, this network is trivial, because the sum of all tissue maps
is the same for all cases, a uniform disk. If, however, this were not the case, training this
network becomes a non-trivial and important step.
The second U-net estimates a calcification support map, where any pixel that contains some
amount of calcium is set to one. The calcium support map is used in the formulation of the SART
algorithm.
The third U-net estimates calcium and adipose tissue maps from the dual-energy FBP images.
The fibroglandular tissue map can be estimated from the difference
of the estimated tissue sum map and the sum of adipose and calcification maps.
With an accurate initial estimate of all tissue maps, team MIR linearized the transmission model
in Eq.~(\ref{transmissionModel}) using the initial estimate as an expansion point.
From this linearized model, they derive a SART algorithm
to refine all of the tissue maps. Team MIR found that convergence of standard SART was slow, so
they used the derived calcification support map to help confine the calcification tissue map
and they made use of a fourth U-net specifically designed to reduce the necessary number of iterations.
They truncated the SART iteration short of convergence and used the intermediate calcification
and adipose map estimates as inputs to the fourth U-net, which then estimates more accurate
calcification and adipose maps. Again, the fibroglandular tissue map is obtained by subtracting
these maps from the sum of tissue maps.
The team MIR members are Hyeongseok Kim (hskim3466@gmail.com) and Seungryong Cho, and both
members are from the
Korea Advanced Institute of Science and Technology (KAIST), Daejeon, South Korea.

The worst-case ROI for this algorithm is shown
in Fig.~\ref{fig:kimhs369-ROIs}. The worst-case ROI is the same one that appeared in the eighth place results
shown in Fig.~\ref{fig:leekunpeng-ROIs} except that the error is confined to a few pixels in the center of
the images and is barely perceptible in the difference images.

\subsubsection*{Second place}

The algorithm developed by team iTORCH is the only one in the top ten performing entries
that did not use CNNs in the data processing. Instead, they used the training data
to derive mathematical constraints on the three tissue maps, which they incorporate
into an iterative image reconstruction algorithm. The basic iterative algorithm handles
the non-linear model for the transmission data in Eq.~(\ref{transmissionModel}) in a manner similar to
the method reported in Chen et al. \cite{chen2021non}. To optimize the
quadratic data fidelity term, the Conjugate Gradients (CG) algorithm is used to obtain
dual-energy images with mitigated artifacts. The tissue maps for adipose and fibroglandular
tissue are then estimated from these images by least-squares. To augment
optimization of the data fidelity term, several regions of the image space are identified
using thresholding of the intermediate maps. First, it is noted that the only variable
part in the image is a circular region just inside the skin-line. This circular region
is separated into "edge" and "non-edge" pixels. For non-edge pixels, tissue maps can only
be 1 or 0, and fractional values are confined to edge pixels. Furthermore, it is also noted
that the calcification map is spatially sparse and that the sum of the tissue maps within
the circular region sum to one.
The full algorithm has an outer iteration, where
the reconstruction region is bipartioned into the above-mentioned regions based on the current
estimates of the tissue maps. Once these regions are identified, the inner iteration
optimizes the data fidelity term with CG, then subsequently enforces the known sum constraint
and the binary value constraint for non-edge pixels. The implementation that achieved
the second place results only used three outer iterations and variable numbers of inner
iterations, [10, 30, 30]. Within each inner iteration CG was run for 200 iterations.
The team iTORCH members are Xiaoyu Hu (Xiaoyu.Hu@UTSouthwestern.edu) and Xun Jia, from the
Department of Radiation Oncology, University of Texas Southwestern Medical Center, Dallas, Texas, USA.
Xun Jia is presently at the Department of Radiation Oncology and Molecular Radiation Sciences,
Johns Hopkins University, Baltimore, Maryland, USA.

The worst-case ROI images are not shown for this algorithm because the maximum pixel discrepancy over all 100 test cases for
all 3 tissue maps is 1.77$\times 10^{-3}$ and this discrepancy would not be seen in the [-0.5,0.5] gray scale
for the difference images.

\subsubsection*{First place}

Team GM\_CNU developed the winning algorithm.
The algorithm is dual-domain joint learning reconstruction method (JLRM) combined with physical
processing for spectral CT (SCT). The overall design of JLRM-SCT includes two main modules:
1) a pre-decomposition module and 2) a main iterative loop.
In each module, both algebraic reconstruction and CNN processing steps are included.

The algebraic  method in the pre-decomposition module uses the extended algebraic
reconstruction technique (E-ART)\cite{zhao2014extended} to reconstruct
two basis images using adipose and fibroglandular tissues as the basis materials.
A CNN is subsequently trained to estimate the
three tissue maps from the two basis images, yielding the pre-decomposition estimate for the
JLRM-SCT algorithm. The particular CNN used is based on RED-CNN.\cite{chen2017low}.

In the iterative loop, the tissue map estimates are used to estimate the dual-energy sinograms by a
physical modeling network. The difference between these sinograms and the actual sinograms are
reconstructed by the standard ART algorithm, yielding residual images. Another CNN is trained that
takes five inputs; two are the residual images, and the other three are the current tissue map estimates
with attention weighting that highlights edges in the tissue maps. From these five inputs, an increment
to the tissue maps is estimated by the trained CNN. This increment is added to the current tissue map
estimates, completing the processing within one iteration. For obtaining the challenge results,
only two iterations of JLRM-SCT are needed. This team found that they obtained the best results
for the CNN in JLRM-SCT loop using residual-to-residual mapping. Also, in the design of their networks,
they noted that in order to avoid overfitting, it is important to use CNNs with a small receptive field,
i.e. small convolution kernels (1 or 3 pixels wide) and avoiding operations that correlate pixels
that are distant from each other.
The team GM\_CNU  members are Genwei Ma (magenwei@126.com) and Xing Zhao, and both are
at the School of Mathematical Sciences, Capital Normal University, Beijing, China. Dr. Zhao
is also affiliated with the Shenzhen National Applied Mathematics Center, Southern University
of Science and Technology, Shenzhen, China.

The worst-case ROI images for the winning entry
are not shown because the maximum pixel discrepancy is an astounding 6.14$\times 10^{-6} \;\;$!

\section{Discussion and conclusion}
\label{sec:discussion}

The accuracy of the top ten algorithms, on predicting the tissue maps of the 100 test cases, range from a mean RMSE
at the level of 1\% to numerically exact (at the level of single precision floating point computation). We also
point out that these impressive results are obtained on an inverse problem in imaging for which there is no
published solution.
Furthermore, the algorithm development was confined to the 10 week span of the competition;
thus it is likely that the performance of many of the algorithms would improve with further study.

The variety of approaches taken is rather striking. The sixth place time, DCPT+Navrit, achieved the best
results for all participants that did not make explicit use of the physical spectral X-ray transmission model.
Their algorithm used only the dual-energy images as input. In doing so, they do implicitly use the
scan geometry information since it is needed for the FBP algorithm that generates the dual-energy images
from dual-energy sinograms. All of the top five algorithms make use of the scan geometry information and the
spectral transmission model. Four of the top five algorithms are a hybrid of CNN-based and physics-motivated
processing. Interestingly, the second place team, iTORCH, did not use deep-learning in any form; their
approach combines known iterative image reconstruction techniques with non-learning-based
image processing. The winning team, GM\_CNU, developed a hybrid iterative/CNN approach and they performed
numerous studies on optimizing the CNN structure for the posed dual-energy CT inverse problem, achieving
truly impressive results.

Many extensions of the DL spectral CT challenge would be interesting for better understanding of the posed
dual-energy CT inverse problem. A major issue for learning-based algorithms is generalizability and this
takes many forms.

First is the fact that the test set is limited to 100 cases, which is necessary for
practical reasons in running the challenge. Mathematically, however, a proposed algorithm for solving
an inverse problem has to work all of the time, and not just on 100 cases. To this point, we refer to
the 2021 DL sparse-view CT challenge\cite{sparseChallengeReport}, which had 4,000 training cases and 100
test cases. The winning team, Robust-and-stable, tried their algorithm, using the same training dataset,
but testing on a much larger 10,000-case dataset in Genzel {\it et al.}; their mean RMSE was 6.4$\times 10^{-6}$;
the main cluster of individual image RMSE results remained below 1$\times 10^{-5}$; and
there was a single extreme outlier at an RMSE of  3$\times 10^{-5}$. For the present DL spectral CT challenge,
the training and test sets includes 1,000 and 100 cases, respectively. It would be of interest to see how
the leading algorithms perform on larger test sets, or, conversely, how do they perform with fewer training cases.

Second, the DL spectral CT challenge uses test phantoms drawn from a single probabistic object model, which
is loosely based on a 2D breast slice. It would be of interest to try different probabilistic breast slice models,
or use models that describe entirely different anatomy with different tissue types.
Assuming that the scan configuration is still fast-kVp switching dual-energy, the
question is how generalizable are the developed algorithms at solving the same inverse problem
with a different object model.

Third, one can consider other forms of spectral CT. There are many possible configurations of dual-energy CT.
One can also add spectral resolutions with, for example, photon-counting detectors. In addition to changing the spectral
resolution, the number of views and view-angle sampling can be varied. These variations in spectral or spatial
sampling also vary the imaging inverse problem itself.
It would be of interest to know how well the developed algorithms can be adapted to changes in the spectral CT
inverse problem.

The 2022 DL spectral CT challenge problem was designed so that it would be necessary to exploit strong prior
information on the scanned objects; there are only two energy windows in the transmission data,
but three tissues in the object model.  Furthermore, the two soft tissues have very similar linear attenuation curves.
The winning team, GM\_CNU, developed a hybrid iterative/deep-learning approach, but the second place team, iTORCH,
also achieved an extremely high degree of accuracy without deep-learning. Team iTORCH developed an
iterative image reconstruction algorithm exploiting ``hand-crafted'' image processing.  Based on the work of the participants,
it seems that the best way to use deep-learning for solving the spectral CT inverse problem is in a hybrid algorithm
where the full knowledge of the physical model is exploited. Still, it is clear that developing such algorithms is not
automatic; despite all of the literature on the use of deep-learning for inverse problems in imaging,
no ``off-the-shelf'' algorithm was used in the challenge. The participants had to work hard to develop their own algorithms
to make accurate predictions of the tissue maps in the test cases.

The DL spectral CT challenge targets a specific formulation of a dual-energy CT inverse problem that has not been
addressed in the literature before. The leading results show highly accurate tissue map predictions that provide
evidence that the specific inverse problem of the challenge can be solved by the developed methods.
We do, however, point out that theoretical
proof of the inverse problem solution seems unlikely, as has been the case in inverse problems for tomographic imaging
in the past couple of decades. This leaves empirical computer simulation studies as the only option to explore this
or related inverse problem solutions. Thus, we look forward to the detailed exposition of the developed algorithms
and their future investigation under various formulations of the spectral CT inverse problem.

\section*{Acknowledgment}
We acknowledge the tremendous assistance provided by the AAPM Working Group on Grand Challenges (WGGC):
Sam Armato,
Kenny Cha,
Karen Drukker,
Keyvan Farahani,
Lubomir  Hadjiyski,
Reshma Munbodh,
Nicholas Petrick, and
Emily Townley.
The support from
Benjamin Bearce and Jayashree Kalpathy-Cramer at the MedICI challenge platform was critical, as they
implemented the challenge website and provided technical support all throughout the challenge period.
We gratefully acknowledge all of the participants of the DL spectral CT challenge.
This work is supported in part by NIH
Grant Nos. R01-EB026282, R01-EB023968, and R21-CA263660.
The contents of this article are solely the responsibility of
the authors and do not necessarily represent the official
views of the National Institutes of Health.

\end{document}